\documentclass[aps,pre,longbibliography,notitlepage,unsortedaddress,floatfix,nofootinbib]{revtex4-1}

\usepackage[format=plain,labelfont={bf,small},textfont=small,justification=raggedright,singlelinecheck=false]{caption}
\usepackage{bm,amsmath,empheq,amssymb,graphicx,stmaryrd,float,subcaption,upgreek,sidecap,dsfont,mathtools,gensymb,multirow,array}
\usepackage[multiple]{footmisc}
\usepackage[abs]{overpic}
\usepackage{hyperref}

\usepackage{comment,footnote,mathrsfs,subcaption,xcolor}

\DeclareMathOperator{\arcsinh}{arcsinh}

\newcommand{\bx}{{\bm x}}
\newcommand{\bX}{{\bf X}}

\newcommand{\jump}[1]{{\left\llbracket #1 \right\rrbracket}}

\newcommand{\bu}{{\bm u}}

\newcommand{\bmo}{{\bm m}}

\newcommand{\bn}{{\bm n}}

\newcommand{\bE}{{\bf E}}

\newcommand{\bF}{{\bf F}}

\newcommand{\bzero}{{\bf 0}}

\newcommand{\bI}{{\bf I}}

\newcommand{\bR}{{\bf R}}

\newcommand{\br}{{\bm r}}
\newcommand{\bd}{{\bm d}}

\usepackage[utf8]{inputenc}
\usepackage[T1]{fontenc}
\usepackage{nomencl}
 \allowdisplaybreaks

\cleardoublepage           
\renewcommand{\nomname}{} 
\setcitestyle{numbers,square}
\graphicspath{{Figures/}}

\begin{document}
\title{A theory of locally impenetrable elastic tubes}
\author{Krishnan Suryanarayanan}
\email{krishnan.s@iitgn.ac.in, krishnan@manit.ac.in}
\affiliation{Indian Institute of Technology Gandhinagar, Palaj, Gujarat-382055}
\author{Harmeet Singh}
\email{harmeet.singh@iitgn.ac.in}
\affiliation{Indian Institute of Technology Gandhinagar, Palaj, Gujarat-382055}
\date{\today}

\begin{abstract}
We present a reduced-order theory for locally impenetrable elastic tubes with uniform circular cross-section. 
The constraint of local impenetrability is incorporated into a variational scheme to derive a complete set of governing equations, jump conditions, and boundary conditions. 
We show that when local impenetrability is actively enforced, the configurations of the tube comprise segments of standard Kirchhoff rod solutions appropriately connected to segments of constant Frenet curvature. 
The theory is illustrated via three examples: a fully flexible tube and an elastic tube, both hanging under self-weight, and a highly twisted elastic tube.
\end{abstract}

\maketitle
	
\section{Introduction}
A fundamental assumption in continuum modeling of material bodies is that matter cannot penetrate itself \cite{antman:nonlinear}.
In other words, no two distinct material points on a body may occupy a common point in ambient space.
While impenetrability is a global restriction on the deformation of material bodies, it is locally implied by the inequality
\begin{align}\label{eq:det_F}
    \det\bF>0\, ,
\end{align}
where $\bF$ is the associated deformation gradient. 
This condition is necessary and sufficient to ensure that no two distinct, infinitesimally close material points overlap upon deformation.
We refer to this restriction on deformation as \emph{local impenetrability}, which is distinct from the constraint of global impenetrability that prevents material points separated by a finite distance on the material manifold from overlapping.
Although all physically admissible solutions must satisfy \eqref{eq:det_F}, it is typically not actively enforced in continuum theories; instead, it usually serves merely as a criterion to discard physically inadmissible configurations.

Our objective in this article is to construct a reduced-order theory of slender elastic tubes of uniform circular cross-section, where the constraint of local impenetrability is actively enforced.
We will refer to such a body as a \emph{locally impenetrable elastic tube}.

Our motivation for this work stems from a broad set of problems concerning slender bodies in which local impenetrability plays an important role \cite{grandgeorge2021,johanns2021,gonzalez2002}.
The theory presented here is built upon the \emph{Kirchhoff rod theory} (KRT) \cite{antman:nonlinear,oreillybook}, which is a widely used model for slender bodies.
KRT entails reducing a slender body to an inextensible material curve, called the base curve, endowed with an orthonormal frame of directors.
The base curve may be interpreted as a curve connecting the centroids of the cross-sections of the slender body.
The director frame is kinematically linked to the base curve by the constraints of inextensibility and unshearability, tracking the rotations of the cross-sections along the rod's length.
The material surrounding the base curve manifests in KRT through specific bending moduli that linearly relate internal moments to bending strains.

In the present theory of locally impenetrable elastic tubes, the material comprising the tube plays an important role beyond simply imparting bending moduli to the base curve.
Under reduced kinematics, local impenetrability of the tube's material places an upper bound on the Frenet curvature of the base curve.
If the base curve is extensible, this upper bound becomes a pointwise function of its stretch and the radius of the cross-section.
For inextensible base curves, as is the case in KRT, the upper bound on the Frenet curvature is uniquely determined by the cross-sectional radius and is thus uniform across the length of the tube.

Local impenetrability in slender bodies is also significant in computing the \emph{ideal shape} of knots \cite{gonzalez2001} and clasps \cite{starostin2003,cantarella2006}.
Loosely speaking, the ideal shape of a knotted curve of a given topology is the configuration it acquires when thickened into the thickest possible smooth tube, subject to the constraint of self-avoidance (implying both local and global impenetrability). 
While the shapes of ideal knots are primarily dominated by the constraint of global impenetrability \cite{gonzalez1999}, it has been observed in \cite{carlen2005} that, up to numerical error, local impenetrability bounds also arise in trefoil knot configurations.
The ideal shape of a clasp is defined as the configuration assumed by two linked filaments of finite thickness pulled against one another.
The shapes of ideal clasps have been computed in \cite{starostin2003,cantarella2006} for the special case where the centerlines of the two filaments lie in mutually orthogonal planes.
Such a configuration referred to as the \emph{ideal orthogonal clasp}.
The constraint of global impenetrability is actively incorporated into computations of the ideal shapes of knots and clasps \cite{johanns2021,grandgeorge2021}.
However, the bound on the Frenet curvature imposed by local impenetrability may still be exceeded in such models (\cite{grandgeorge2021}, Figure S16), rendering the resulting configurations non-smooth and unphysical.
The current theory could therefore be of relevance in computing both globally and locally impenetrable ideal shapes of knots and clasps.

In the present work, we derive a theory of locally impenetrable elastic rods via a variational framework.
The upper bound on the Frenet curvature of the base curve, implied by the impenetrability constraint \eqref{eq:det_F}, is converted to an equality constraint by introducing a slack function. 
The associated functional is then augmented with this equality constraint using a Lagrange multiplier.
Subsequent derivation of the corresponding Euler-Lagrange equations shows that a deformed configuration of an elastic tube comprises solutions of standard KRT connected by regions of constant Frenet curvature.
We refer to the former regions as \emph{inactive} and the latter as \emph{active}.
An active region nucleates at a point where the curvature of the rod reaches its upper bound and spreads as the rod is further deformed.
The active and inactive regions are connected by appropriate jump conditions obtained naturally from the variational framework.
The location and the extent of the active region are unknowns of the problem, determined as part of the system solution.
The presence of the local impenetrability constraint also modifies the constitutive relations within the active regions: whereas inactive regions are characterized by a linear constitutive relation between internal moment and bending strain, active regions manifest a distinctly nonlinear relationship between these quantities.

We illustrate our theory with three representative examples.
We first compute planar configurations of a locally impenetrable, fully flexible tube of a given thickness hanging under self-weight, with its terminal ends separated by a horizontal distance.
When the two ends of the flexible tube are brought sufficiently close, an active region nucleates and spreads at the lowest point of the tube.
We show that, in contrast to inactive regions, local impenetrability in the active zone produces non-zero internal moments and shear forces.
In the second example, we add bending elasticity to the tube and treat it as an elastic rod under self-weight.
We identify a scaling parameter that measures the relative strength of gravity against the bending stiffness of the elastic tube.
It is shown that, above a critical value of this scaling parameter, the local impenetrability constraint is activated when the ends are sufficiently close.
For smaller values of the parameter, bending elasticity inherently ensures that the deformation never violates local impenetrability.

Finally, we compute highly twisted 3D solutions of an elastic tube using our theory.
In the absence of local impenetrability, the bending energy of the tube tends to concentrate away from its terminal ends.
With local impenetrability incorporated, a small active region nucleates at the mid-length of the rod in a helical form, gradually spreading across the length of the rod as the twist is increased.

Prior works on the subject primarily deal with formulating mathematical conditions that simultaneously enforce both local and global impenetrability constraints.
While the present work deals only with incorporating local impenetrability, it is relevant to mention the following three works that have been instrumental in developing a cogent mathematical description of self-contact in elastic rods.
The first is by Gonzalez and Maddocks \cite{gonzalez1999}, who introduced the notion of the global radius of curvature. 
This geometric framework simultaneously accounts for both local and global self-contact in the characterization of ideal knot shapes. 
The second is the subsequent work by Gonzalez et al. \cite{gonzalez2002}, which proves that despite the inherently non-smooth nature of self-contact in knots, the difference quotient for the second derivative of the centerline remains bounded, and the second derivative of the centerline exists almost everywhere. 
However, it may admit discontinuities at specific points.
The results presented in Sec.~\ref{sec:ideal_tube}, \ref{sec:elastica_under_gravity}, and \ref{sec:3D_sol} below, where only local contact is active, are fully compatible with these regularity conditions. 
Third, Schuricht and von der Mosel \cite{schuricht2003} derived the exact Euler-Lagrange equations for an elastic rod, incorporating both local and global impenetrability via the global radius of curvature \cite{gonzalez1999} while accounting for topological constraints (i.e., knot type) in a variational framework. 
The governing equations they presented are integro-differential in nature.
Furthermore, they rigorously demonstrated that the contact forces generated by self-contact act purely normal to the centerline surface, a property assumed a priori in the wider literature.

The rest of the manuscript is organized as follows.
We begin in Sec.~\ref{sec:impenetrability} by reducing constraint \eqref{eq:det_F} along the base curve of a tubular volume.
In Sec.~\ref{sec:governing_equations}, the variational formulation of the theory is presented.
The force and moment balance equations, as well as the effective relations between internal moments and bending strains in both active and inactive regions, are derived.
Various integrals of the present theory are discussed and compared with their counterparts in standard Kirchhoff rod theory.
We show that integrals associated with the standard KRT persist in the presence of the local impenetrability constraint.
In Sec.~\ref{sec:ideal_tube}, configurations of a locally impenetrable, fully flexible tube hanging under self-weight are computed.
Sec.~\ref{sec:elastica_under_gravity} contains solutions for a locally impenetrable elastic tube hanging under self-weight.
3D solutions for a highly twisted locally impenetrable elastic tube are presented in Sec.~\ref{sec:3D_sol}.
We conclude in Sec.~\ref{sec:conclusion} with a discussion of our results and potential future extensions.
The details of the variational procedure employed in Sec.~\ref{sec:governing_equations} are set forth in Appendix \ref{app:variational_derivation}.

\section{Impenetrability of a tubular volume}\label{sec:impenetrability}
\begin{figure}[t]
    \centering
    \includegraphics[trim={4.5cm 3cm 8cm 3cm}, clip,width=0.75\linewidth]{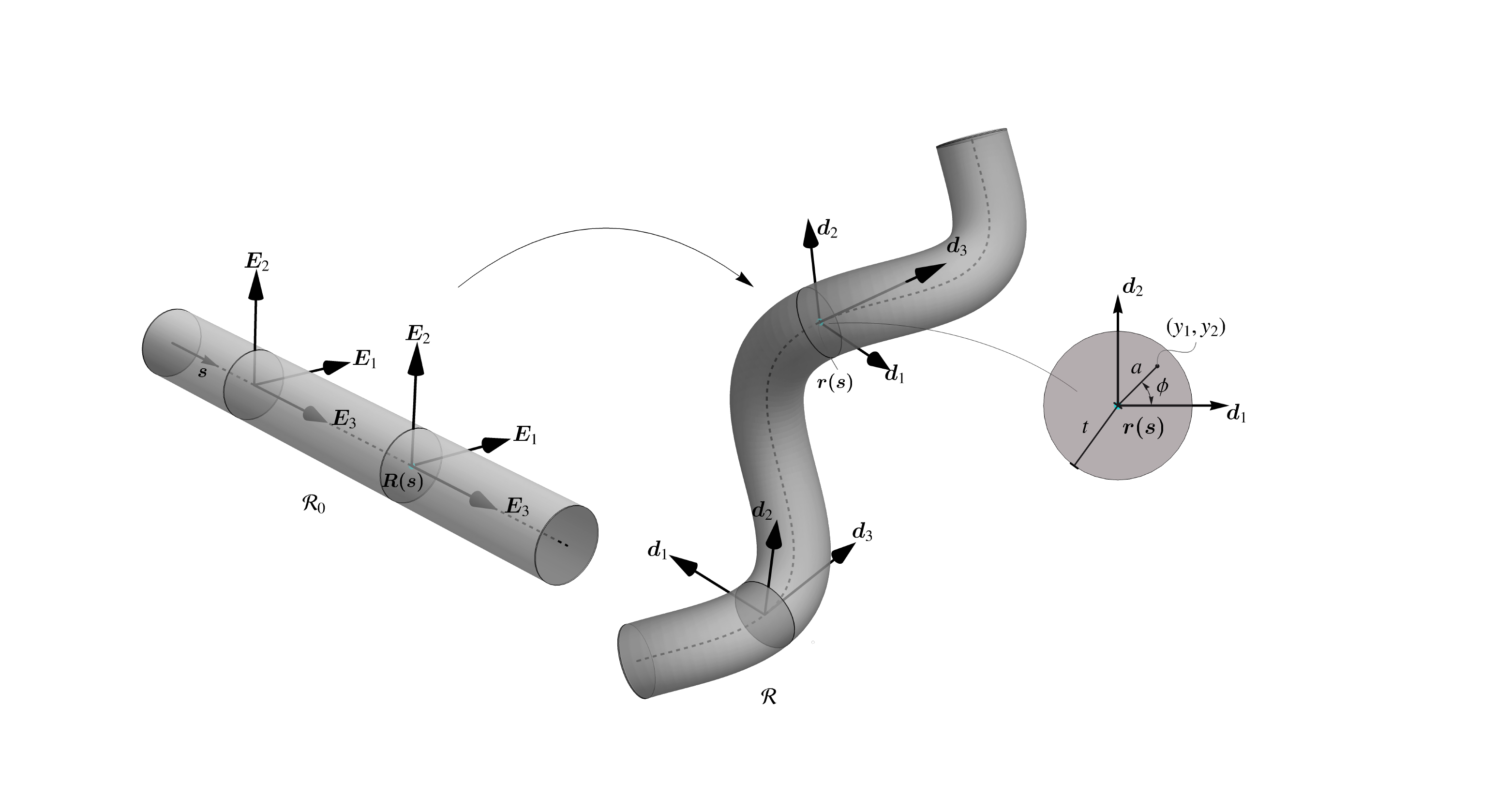}
    \caption{(Left) Reference configuration $\mathcal{R}_0$ of a circular tube of radius $t$ and centerline $\bR(s)$ parameterized by the arc length $s$. (Middle) Deformed configuration $\mathcal{R}$ with centerline $\br(s)$ and the associated orthonormal director frame $\{\bd_1,\bd_2,\bd_3\}$. (Right) Cross-sectional geometry of the tube.}
    \label{fig:schematic}
\end{figure}
Consider a slender body $\mathcal{B}$ whose stress-free reference configuration $\mathcal{R}_0$ is a straight tube\footnote{We choose to work with a straight reference configuration for simplicity. The kinematics of a curved reference configuration do not fundamentally differ from those of a straight rod.} with an arbitrary cross-section that may vary along its length.
To each point $s$ on $\bR$, we attach a right-handed, orthonormal, constant frame of directors $\bE_i$, $i\in\{1,2,3\}$, such that $\bE_3$ coincides with the tangent, i.e.,
\begin{align}\label{eq:reference_tangent_constraint}
   \bR'(s) = \bE_3\, .
\end{align}
Here, $()'$ denotes the derivative w.r.t. $s$.
We parametrize $\mathcal{R}_0$ as
\begin{align}\label{eq:reference_config_parametrization}
    \bX(s,y_1,y_2) = \bR(s) + y_1 \bE_1 + y_2 \bE_2\, ,
\end{align}
where $\bX\equiv\bX(s,y_1,y_2)$ denotes the position vector of a material point, identified by the coordinates $\{s,y_1,y_2\}\in\mathcal{B}$, in $\mathcal{R}_0$.

Let $\mathcal{B}$ deform into a non-straight tube and acquire a configuration $\mathcal{R}$.
The centerline curve $\bR\in\mathcal{R}_0$ convects into $\br\equiv\br(s)\in\mathcal{R}$ with the deformation.
We endow $\br$ with an orthonormal frame of directors $\bd_i\equiv\bd_i(s)$, $i\in\{1,2,3\}$, such that $\bd_3(s)$ aligns with the tangent at $\br$ for all $s$, 
\begin{align}\label{eq:current_config_director_alignment}
\br'(s) = v_3(s) \bd_3(s)\, ,
\end{align}
where $v_3\equiv v_3(s)$, defined as $v_3(s):=|\br'(s)|$, is the stretch of the centerline.
In choosing the above representation, we have assumed the deformations of the rod to be unshearable.
The orthonormality of the director frame requires that each director evolves along $s$ as
\begin{align}\label{eq:director_orthonormality}
    \bd'_i(s) = \bu(s)\times\bd_i(s)\, , 
\end{align}
where $\bu\equiv\bu(s)$ is the Darboux vector associated with the director frame.
We parametrize $\mathcal{R}$ as
\begin{align}\label{eq:current_config_parametrization}
\bx(s,y_1,y_2) = \br(s) + y_1\bd_1(s) + y_2\bd_2(s)\, ,
\end{align}
where $\bx\equiv\bx(s,y_1,y_2)$ denotes the current position vector of material points in $\mathcal{R}$.

Using \eqref{eq:reference_config_parametrization} and $\eqref{eq:current_config_parametrization}$, the deformation gradient $\bF\equiv\bF(\bX)$ at any material point can be computed as
\begin{align}\label{eq:deformation_gradient}
    \bF = \bd_1\otimes\bE_1 + \bd_2\otimes\bE_2 + \left[\br' + (-y_1 u_2 + y_2 u_1)\bd_3 + y_1 u_3\bd_2 - y_2u_3\bd_1\right]\otimes\bE_3\, ,
\end{align}
where $u_i=\bu\cdot\bd_i$.
Furthermore, it follows from \eqref{eq:deformation_gradient} that
\begin{align}\label{eq:determinant_F}
    \det \bF = v_3-y_1 u_2 + y_2 u_1\, .
\end{align}
Local impenetrability of matter imposes $\det\bF>0$, therefore,
\begin{align}\label{eq:determinant_F_positive}
    v_3-y_1u_2 + y_2 u_1>0\, .
\end{align}
The inequality above must hold for all values of $\{s,y_1,y_2\}$ in the tubular volume. 
Choosing $y_1=y_2=0$, we conclude that $v_3>0$.

Let us represent the point $\{y_1,y_2\}$ on the cross-section in polar coordinates $\{a,\phi\}$ so that $y_1 = a\cos\phi$ and $y_2 = a\sin\phi$.
Inequality \eqref{eq:determinant_F_positive} then takes the form
\begin{align}\label{eq:determinant_F_in_polar}
    v_3 + a\sqrt{u_1^2 + u_2^2}\sin(\phi + \alpha)>0\, ,\qquad\text{where}\qquad \tan\alpha = -\frac{u_2}{u_1}\, .
\end{align}
Substituting $\kappa = \sqrt{u_1^2+u_2^2}$, where $\kappa\equiv\kappa(s)$ is the Frenet curvature of the centerline $\br$ \cite{oreillybook}, and rearranging yields
\begin{align}\label{eq:kappa_inequality_0}
    \sin(\phi+\alpha)>-\frac{v_3}{\kappa a}\, .
\end{align}
To ensure this inequality holds for all values of $\phi$, the right-hand side must be strictly less than the minimum value of the sine function ($-1$), which requires
\begin{align}\label{eq:kappa_inequality_1}
    \frac{v_3}{\kappa a}>1.
\end{align}

We now assume for the rest of the article that the tube under consideration has a uniform circular cross-section of radius $t$, so that $0\le a\le t$ (Fig~\ref{fig:schematic}).
Inequality \eqref{eq:kappa_inequality_1} is then guaranteed to hold for all values of $a$ if $\kappa<\tfrac{v_3}{t}$, transferring the local impenetrability constraint to the centerline of the tube.
In the following, we will only concern ourselves with Kirchhoff rods, whose kinematics dictate that the centerline be unstretchable, i.e., $v_3 = 1$.
Therefore, the impenetrability constraint for an elastic tube of circular cross-section reduces to
\begin{align}\label{eq:kappa_inequality_strong}
    \kappa<\frac{1}{t}\, .
\end{align}

It could be argued that the assumptions of the inextensibility of the centerline and the unshearability of the cross-sections are valid only in a regime where $\kappa t \ll 1$.
Consequently, enforcing these kinematic constraints in a limit where $\kappa t \approx 1$ may be deemed unreasonable.
However, it is well known that Kirchhoff rod theory gives surprisingly good results even when pushed beyond the assumptions upon which it is built. 
A particular example is the remarkable agreement between certain predictions made by a Kirchhoff rod model and experimental results pertaining to an orthogonal clasp \cite{grandgeorge2021}.
Our choice to construct our theory using the kinematics of KRT is motivated by this observation. 

The entire procedure outlined in this section can be modified rather easily to include richer kinematics (such as non-straight reference configurations, extensibility, shear deformations, etc.) to obtain a modified version of \eqref{eq:kappa_inequality_strong}.
In the same spirit, the variational procedure outlined below has been kept sufficiently general by not assuming any particular form of the energy density for the tube.

\section{Equations governing a locally impenetrable Kirchhoff rod}\label{sec:governing_equations}
For the sake of simplicity, we will work with the following weakened form of the inequality constraint \eqref{eq:kappa_inequality_strong}:
\begin{align}\label{eq:kappa_inequality}
    \kappa\le\frac{1}{t}\, ,
\end{align}
where we view the saturation of the above inequality as equivalent to $\lim_{\epsilon\rightarrow 0}\left(\kappa - \frac{1}{t} + \epsilon\right) = 0$, with $\epsilon>0$. 

The kinematics of a Kirchhoff rod remain identical to \eqref{eq:current_config_director_alignment} and \eqref{eq:director_orthonormality}, except that the centerline is assumed to be unstretchable, i.e., $v_3 = 1$.
Consequently, equation \eqref{eq:current_config_director_alignment} modifies to
\begin{align}\label{eq:kirchhoff_kinematic}
    \br'(s)=\bd_3(s)\, .
\end{align}

We will adopt a variational scheme to obtain all the necessary equations governing a locally impenetrable elastic tube.
We begin by converting the inequality constraint \eqref{eq:kappa_inequality} to an equality constraint by means of a slack function $g\equiv g(s)$,
\begin{align}\label{eq:kappa_equality}
    \kappa(s) - \hat\kappa + g(s)^2 = 0\, ,
\end{align}
where we use $\hat\kappa \equiv 1/t$ for brevity of notation.
We posit that static equilibria of a locally impenetrable elastic tube are extrema of the following action functional:
\begin{align}\label{eq:action_functional}
    \mathcal{A}[\br,\bd_i,g] = \int_0^L\mathcal{L}\left(\br(s),\br'(s),\bd_i(s),u_i(s),\kappa(s),g(s)\right)ds\, ,
\end{align}
where the Lagrangian density $\mathcal{L}$ is given by
\begin{align}\label{eq:lagrangian_density}
    \mathcal{L} = \mathcal{W} + \bn\cdot\left(\br' - \bd_3\right) + \Lambda\left[\kappa - \hat\kappa + g^2\right] - \mathcal{V}\, .
\end{align}
Here, $\mathcal{W}\equiv\mathcal{W}(u_i)$ is an elastic energy density (per unit length) function\footnote{Assuming the existence of $\mathcal{W}(u_i)$ restricts the validity of our analysis to hyperelastic rods only.}, assumed to be strictly convex in its arguments and minimized at $u_i = 0$\footnote{A non-straight intrinsic shape of the rod, described by $\hat{u}_i\equiv \hat{u}_i(s)$, can be accounted for by considering an energy density $\mathcal{W}\equiv\mathcal{W}(u_i-\hat{u}_i)$ that is minimized at $u_i=\hat{u}_i$\cite{antman:nonlinear,dichmann1996}.}; $\bn\equiv\bn(s)$ is a vector Lagrange multiplier function that enforces constraint \eqref{eq:kirchhoff_kinematic}, $\Lambda\equiv\Lambda(s)$ is a scalar Lagrange multiplier that enforces \eqref{eq:kappa_equality}, and $\mathcal{V}\equiv\mathcal{V}(\br)$ is a potential function dependent on $\br$ alone.
The function $\mathcal{V}$ can be used to incorporate position-dependent potentials, such as the gravitational potential.

We subject \eqref{eq:action_functional} to the following infinitesimal variations:
\begin{align}\label{eq:variations}
s^*:=s+\delta s\, ,\qquad \br^*(s^*):=\br(s) + \delta\br(s)\, , \qquad \bd^*_i(s^*):=\bd_i(s) + \delta\bd_i(s)\, ,\qquad g^*(s^*):=g(s) + \delta g(s)\, .
\end{align}
The variation in the arc-length coordinate $s$ is to be interpreted as a shift in the material \emph{label} of a material \emph{point}.
The variational operator $\delta$ is to be interpreted as denoting an infinitesimal change in a field at a \emph{fixed material point}.
For instance, the field $\br^*(s^*)$ denotes the perturbed position vector of a material point whose label prior to the perturbation $\delta$ was $s$, and whose new label is $s^*$, the underlying material point remaining the same.

The variation in \eqref{eq:action_functional}, up to linear order, induced by \eqref{eq:variations} is given by
\begin{align}\label{eq:variation_A_1}
    \delta \mathcal{A} = &\int_0^L\left[\bn\cdot\delta\br + \bmo\cdot\delta\bm{\theta} - H \delta s\right]' ds - \int_0^L \left(\bn' + \frac{\partial\mathcal{V}}{\partial\br}\right)\cdot\delta\br ds - \int_0^L\left(\bmo' + \bd_3\times\bn\right)\cdot\delta\bm{\theta} ds \nonumber\\
    &+ \int_0^L\left\{\left(\bn' + \frac{\partial\mathcal{V}}{\partial\br}\right)\cdot\br' + \left(\bmo' + \bd_3\times\bn\right)\cdot\bu - 2\Lambda g g' \right\}\delta s ds + \int_0^L2\Lambda g \delta g ds\, ,
\end{align}
where we define $\bmo\equiv\bmo(s)$ and $H\equiv H(s)$ as
\begin{align}
    \bmo &:= \frac{\partial\mathcal{W}}{\partial u_i}\bd_i + \frac{\Lambda}{\kappa}\mathbb{P}\bu\, ,\label{eq:constitutive_relation}\\
    H &:= \bn\cdot\br' + \bmo\cdot\bu - \left(\mathcal{W} - \mathcal{V}\right)\, ,\label{eq:Hamiltonian}
\end{align}
where $\mathbb{P} = \bI - \bd_3\otimes\bd_3$, $\bI$ being the identity tensor.
The details of the calculation leading to \eqref{eq:variation_A_1} are postponed to Appendix \ref{app:variational_derivation}.

Let there exist a corner point $s_0\in[0,L]$ \cite{gelfandfomin2000} where there may exist a jump discontinuity in $\bn$, $\bmo$, $\bu$, $\mathcal{V}$, and consequently in $H$, with the position and the tangent vector remaining continuous: 
\begin{align}\label{eq:position_tangent_jump_conditions}
    \jump{\br} = \bzero\, ,\qquad \jump{\br'} = \bzero\, .
\end{align}
Here, $\jump{A} = A(s^+) - A(s^-)$, where $A(s^\pm) = \lim_{\epsilon\rightarrow 0}A(s\pm\epsilon)$ with $\epsilon>0$, represents the jump in a field $A$ across $s_0$.
In light of that, the first integral in \eqref{eq:variation_A_1} can be evaluated to obtain the following:
\begin{align}\label{eq:variation_final}
    \delta \mathcal{A} = &\bn\cdot\delta\br\big|_0^L - \jump{\bn}\cdot\delta\br(s_0) + \bmo\cdot\delta\bm{\theta}\big|_0^L - \jump{\bmo}\cdot\delta\bm{\theta}(s_0) - H\delta s\big|_0^L + \jump{H}\delta s_0\nonumber\\
     &- \int_0^L \left(\bn' + \frac{\partial\mathcal{V}}{\partial\br}\right)\cdot\delta\br ds - \int_0^L\left(\bmo' + \bd_3\times\bn\right)\cdot\delta\bm{\theta} ds\nonumber\\
    &+ \int_0^L\left\{\left(\bn' + \frac{\partial\mathcal{V}}{\partial\br}\right)\cdot\br' + \left(\bmo' + \bd_3\times\bn\right)\cdot\bu - 2\Lambda g g' \right\}\delta s ds + \int_0^L2\Lambda g \delta g ds.
\end{align}
We now proceed systematically to obtain the conditions that render $\delta \mathcal{A} = 0$.

\subsection{Inactive and active regions}
Let us first consider a case where $\{\delta s, \delta\br,{\delta\bm \theta}\} = \{0,\bzero,\bzero\}$.
Then, $\delta \mathcal{A} = 0$ implies
\begin{align}\label{eq:lambda_g_0}
    \Lambda g = 0\, ,
\end{align}
meaning that in any subregion of $[0,L]$, either $\Lambda$ or $g$ must vanish.
Regions with $\Lambda=0$ will be referred to as \emph{inactive}, indicating that the inequality constraint \eqref{eq:kappa_inequality} is inactive.
Similarly, regions with $g=0$ will be called \emph{active}.
For the sake of simplicity, and without loss of generality, we associate the interval $(0,s_0)$ with the inactive region, and $(s_0,L)$ with the active region in $[0,L]$.
From \eqref{eq:kappa_equality} and \eqref{eq:lambda_g_0}, the following conditions on the Frenet curvature of an elastic rod are established:
\begin{subequations}\label{eq:active_inactive_regions}
\begin{align}
\Lambda &=0\, ,\qquad \kappa<\hat\kappa\, ,\qquad s\in(0,s_0)\, ,\qquad\text{\emph{inactive region}}\, , \label{eq:inactive_region}\\
g &= 0\, ,\qquad \kappa=\hat\kappa\, ,\qquad s\in(s_0,L)\, ,\qquad \text{\emph{active region}}\, .\label{eq:active_region}
\end{align}
\end{subequations}

In general, under complex loading, a Kirchhoff rod may develop multiple active regions within its length, separated from the inactive regions by boundaries whose locations are not known \emph{a priori}.
Computing the locations of these unknown boundaries requires a jump condition on the Hamiltonian function, which is derived in Sec.~\ref{sec:conservation_laws}.

\subsection{Force and moment balance}
The force and moment balance equations valid in $[0,L]\setminus\{s_0\}$ are obtained as an implication of $\delta \mathcal{A} = 0$ in \eqref{eq:variation_final} by considering variations $\delta\br$ and $\delta\bm\theta$ that vanish only at $s=\{0,s_0,L\}$, with $\delta s=0$ for all $s$.
The resulting conditions are
\begin{subequations}\label{eq:force_moment_balance}
\begin{align}
\bn' + \frac{\partial\mathcal{V}}{\partial\br} &=\bzero\, ,\qquad \forall s\in(0,L)\setminus\{s_0\}\, ,\label{eq:force_balance}
\\
\bmo' + \br'\times\bn &= \bzero\,,\qquad \forall s\in(0,L)\setminus\{s_0\}\, .\label{eq:moment_balance}
\end{align}
\end{subequations}
By comparison with the standard force and moment balance laws of a Kirchhoff rod, we identify the Lagrange multiplier $\bn$ as the internal force and $\bmo$ as the internal moment in the rod.
Both equations hold as written in the inactive and active regions of the rod.

The force and moment balance conditions valid at the interface $s=s_0$ are obtained by considering $\delta\br$ and $\delta\bm\theta$ to vanish everywhere on the domain except at $s=s_0$.
Then, $\delta \mathcal{A}=0$ in \eqref{eq:variation_final} implies
\begin{subequations}\label{eq:force_moment_jump_conditions}
    \begin{align}
    \jump{\bn} &= \bzero\, ,\qquad\text{at}\qquad s=s_0\, ,\label{eq:force_jump_condition}
    \\
    \jump{\bmo} &= \bzero\, ,\qquad\text{at}\qquad s=s_0\, .\label{eq:moment_jump_condition}
\end{align}
\end{subequations}
In other words, the internal force and moment must be continuous across any interface separating the inactive and active regions.

\subsection{Constitutive relations}
The constitutive relations for a locally impenetrable Kirchhoff rod can be written as a piecewise function using \eqref{eq:constitutive_relation} and \eqref{eq:active_inactive_regions}:
\begin{align}\label{eq:piecewise_constitutive_relation}
\bmo = 
    \begin{dcases}
        \frac{\partial\mathcal{W}}{\partial u_i}\bd_i \quad &\text{for} \quad s\in(0,s_0)\\
        \frac{\partial\mathcal{W}}{\partial u_i}\bd_i + \frac{\Lambda}{\hat\kappa}\mathbb{P}\bu \quad &\text{for} \quad s\in(s_0,L)\, .
    \end{dcases}
\end{align}
To reiterate, $(0,s_0)$ is the inactive region, whereas $(s_0,L)$ is the active region.
The constitutive relation in the inactive region is the well-known constitutive relation for a Kirchhoff elastic rod \cite{audoly2016,oreillybook}.
In the active region, the constitutive relation contains an additional term that warrants further discussion.

It can be shown that the vector $\mathbb{P}\bu/\hat\kappa$ is of unit norm and is identical to the Frenet binormal of $\br(s)$.
The following properties of this vector will prove useful later:
\begin{align}\label{eq:PU_orthonormality}
    \mathbb{P}\bu\cdot\bd_3 = 0\, ,\qquad\mathbb{P}\bu\cdot\bd'_3 = 0\, ,
    \qquad\mathbb{P}\bu\cdot\bu' = 0\, ,
\end{align}
where the first two conditions are identities, whereas the third condition results from using $u_1u'_1+u_2 u'_2=0$, obtained by differentiating the relation $u_1^2+u_2^2 = \hat\kappa^2$.

In Kirchhoff rod theory, the energy density function takes the following form:
\begin{align}\label{eq:Kirchhoff_energy_density}
    \mathcal{W} = \sum_{i=1}^{3}\frac{1}{2}K_i u_i^2\,\quad\text{for} \quad s\in(0,L)\setminus\{s_0\}\, ,
\end{align}
where $K_i$, $i\in\{1,2,3\}$, are the bending moduli of the rod about $\bd_i$.
Since we are only concerned with elastic tubes of circular cross-section, we have $K_1=K_2=K$.
The following relations follow from \eqref{eq:piecewise_constitutive_relation} and \eqref{eq:Kirchhoff_energy_density}:
\begin{align}\label{eq:constitutive_law_inactive}
    m_1 = K u_1\,,\quad m_2 = K u_2\,,\quad m_3 = K_3 u_3\, ,\qquad \text{for} \qquad s\in(0,s_0)\, ,
\end{align}
where $m_i\equiv\bmo\cdot\bd_i$, $i\in\{1,2,3\}$.
Thus, the director components of the internal moments are linear functions of the respective director components of the Darboux vector in the inactive region.

The constitutive relations in the active region can be obtained by using \eqref{eq:Kirchhoff_energy_density} in \eqref{eq:piecewise_constitutive_relation} and rearranging to yield
\begin{align}\label{eq:constitutive_law_active}
    m_1 = \left(K+\frac{\Lambda}{\hat\kappa}\right) u_1\, ,\qquad m_2 = \left(K+\frac{\Lambda}{\hat\kappa}\right)u_2\, ,\qquad m_3 = K_3 u_3\,, \qquad \text{for} \qquad s\in(s_0,L)\, .
\end{align}
The first two relations represent a notable departure from the standard constitutive relations of Kirchhoff rod theory:
the internal moments $m_1$ and $m_2$ are no longer linear functions of $u_1$ and $u_2$ due to the presence of the function $\Lambda(s)$. 
The term in parentheses can be interpreted as a variable stiffness measure of the rod, which adjusts dynamically to ensure that the constraint of constant Frenet curvature (i.e., $u_1^2 + u_2^2 = \hat\kappa^2$) is maintained in the active region.
This constraint, in conjunction with the expressions for $m_1$ and $m_2$ in \eqref{eq:constitutive_law_active}, can be used to express $\Lambda$ explicitly in terms of the moments as
\begin{align}\label{eq:Lambda}
    \Lambda = \left(m_1^2 + m_2^2\right)^{1/2} - K \hat\kappa\, .
\end{align}
Substituting \eqref{eq:Lambda} into \eqref{eq:constitutive_law_active}, the following nonlinear relations between the bending strains and the internal moments in the active region are established:
\begin{align}\label{eq:nonlinear_constitutive_relation}
    u_1 = \frac{m_1\hat\kappa}{\left(m_1^2 + m_2^2\right)^{1/2}}\, ,\qquad u_2 = \frac{m_2\hat\kappa}{\left(m_1^2 + m_2^2\right)^{1/2}}\, ,\qquad u_3 = \frac{m_3}{K_3}\, ,\qquad \text{for} \qquad s\in(s_0,L)\, .
\end{align}

\subsection{Integrals}\label{sec:conservation_laws}
We discuss here the effect of the impenetrability constraint \eqref{eq:kappa_inequality} on the first integrals that standard Kirchhoff rod theory admits \cite{steigmannfaulkner1993,maddocksdichman1994}.
We show that all the standard first integrals persist despite the modified constitutive relation \eqref{eq:piecewise_constitutive_relation}.

The first integrals discussed below can all be obtained from \eqref{eq:variation_final} using Noether's theorem \cite{hill1951,singh2019,singh2021,neukirch2025}.
In order to keep the discussion brief, we choose to present shorter alternative derivations instead.

The first integrals for force and moment balance can be read directly from \eqref{eq:force_moment_balance}.
If $\partial\mathcal{V}/\partial\br=\bzero$, then \eqref{eq:force_moment_balance} can be integrated in both the inactive and active regions to yield two conserved quantities.
In addition, the continuity of the internal force and moment \eqref{eq:force_moment_jump_conditions} at $s_0$ ensures that these two conserved vector fields assume the same value throughout $(0,L)$.
Therefore, we have
\begin{subequations}\label{eq:force_moment_conservation}
\begin{alignat}{2}
    \bn(s) &= \bn(0)\, ,&s\in(0,L)\, ,\label{eq:force_conservation}
    \\
    \bmo(s) + \br(s)\times\bn(s) &= \bmo(0) + \br(0)\times\bn(0)\, ,  \qquad &s \in(0,L)\, .\label{eq:moment_conservation}
\end{alignat}
\end{subequations}
These two first integrals are independent of the constitutive relations of the elastic rod and are therefore, unsurprisingly, agnostic to \eqref{eq:piecewise_constitutive_relation}.

Next, we take the inner product of the moment balance \eqref{eq:moment_balance} with $\bd_3$, use \eqref{eq:kirchhoff_kinematic}, and rearrange the resulting equation as $(\bmo\cdot\bd_3)' - \bmo\cdot\bd'_3=0$.
Using \eqref{eq:director_orthonormality} and \eqref{eq:constitutive_law_inactive}, it can be shown that the twist $u_3$ is conserved in the inactive region.
The same is implied in the active region by the constitutive laws \eqref{eq:constitutive_law_active} and \eqref{eq:PU_orthonormality}$_{1,2}$.
This, together with the continuity of the internal moment \eqref{eq:moment_jump_condition}, implies
\begin{align}\label{eq:twist_conservation}
    u_3(s) = u_3(0)\, ,\qquad \text{for} \qquad s\in(0,L)\, .
\end{align}

The final conservation law is obtained by summing the inner product of the force balance \eqref{eq:force_balance} with $\br'$ and the moment balance \eqref{eq:moment_balance} with $\bu$, respectively.
Rearranging the resulting expression using \eqref{eq:piecewise_constitutive_relation} and \eqref{eq:PU_orthonormality}, it can be shown that the Hamiltonian function $H(s)$ is conserved in both the inactive and active regions.
Furthermore, setting $\delta \mathcal{A} = 0$ for a variation in the boundary position $\delta s_0$ while keeping all other variations in \eqref{eq:variations} equal to zero implies $\jump{H}\delta s_0 = 0$. Since $\delta s_0$ is arbitrary, this yields
\begin{align}\label{eq:Hamiltonian_continuity}
    \jump{H}=0\, .
\end{align}
In view of the above, the following conservation law holds throughout, and across, the inactive and active regions:
\begin{align}\label{eq:Hamiltonian_conservation}
    H(s)=H(0)\, ,\qquad s\in(0,L)\, .
\end{align}

In the next three sections, we use our theory to formulate and solve three specific examples that incorporate the impenetrability constraint \eqref{eq:kappa_inequality}.

\section{A Fully flexible tube hanging under self-weight}\label{sec:ideal_tube}
A fully flexible string (i.e., one with no bending elasticity, or $\bmo = \bzero$) hanging in a uniform gravitational field under its own weight takes a planar shape called the \emph{catenary} \cite{timoshenko1965}.
As the two ends of the string are brought closer together, the curvature at its lowest point tends to infinity.
Here, we use the present theory to compute the equilibrium shapes of a fully flexible, locally impenetrable tube hanging under its self-weight.
From this point onward, we will refer to such a structure simply as a flexible tube.

Consider a flexible tube of radius $t$ and length $L$ hanging under self-weight with its two ends separated horizontally by a distance $\Delta$. 
The lowest point on the centerline of the tube attains the maximum permissible curvature $\hat\kappa = 1/t$ when $\Delta=(2L/\hat\kappa)\arcsinh(\hat\kappa/2)$.
If $\Delta$ is reduced any further, an active region of constant curvature $\hat\kappa$ nucleates at the lowest point and propagates.
Let us identify this active region with the unknown interval $(s_1,s_2)$.

We choose the plane of deformation of the tube such that it is spanned by $\{\bd_1,\bd_3\}$, and we pick a Cartesian basis $\bE_i$, $i\in\{1,2,3\}$, so that $\bE_2=\bd_2$.
The directors $\{\bd_1,\bd_3\}$ differ from $\{\bE_1,\bE_3\}$ by an anticlockwise rotation of magnitude $\theta$, meaning that the various fields of interest admit the following representations:
\begin{subequations}\label{eq:planar_representations}
    \begin{alignat}{3}
    & \br = r_1\bE_1 + r_3\bE_3\, ,\qquad && \bd_1 = \cos\theta\bE_1 - \sin\theta\bE_3\, ,\qquad &&\bd_3 = \sin\theta\bE_1 + \cos\theta\bE_3\, ,\label{eq:planar_representation_kinematic}
    \\
    & \bn = n_1\bd_1 + n_3\bd_3\, ,\qquad && \bmo = m_2 \bd_2\, ,\qquad && H = n_3 + \frac{m_2^2}{2K}\, .\label{eq:planar_representation_mechanics}
\end{alignat}
\end{subequations}

The kinematics of a planar material curve $\{\br,\bd_i\}$, $i\in\{1,3\}$, are governed by relations \eqref{eq:kirchhoff_kinematic} and \eqref{eq:director_orthonormality}.
Substituting \eqref{eq:planar_representation_kinematic} into \eqref{eq:kirchhoff_kinematic} and \eqref{eq:director_orthonormality}, the following scalar relations are obtained:
\begin{align}\label{eq:2D_kinematics}
    r'_1 = \sin\theta\, ,\qquad r'_3 = \cos\theta\, ,\qquad \theta' = u_2\, ,\qquad \forall s\in[0,L]\, .
\end{align}

A fully flexible, inextensible, and unshearable tube has no elastic energy associated with it, i.e., $\mathcal{W} = 0$.
Therefore, from \eqref{eq:piecewise_constitutive_relation}, the constitutive relation for an impenetrable ideal tube reduces to
\begin{align}\label{eq:piecewise_constitutive_relation_string}
\bmo = 
    \begin{dcases}
        \bzero \quad &\text{for} \quad s\in(0,L)\setminus(s_1,s_2)\\
         \Lambda\bd_2\quad &\text{for} \quad s\in(s_1,s_2)\, ,
    \end{dcases}
\end{align}
implying that the internal moment in the active region—unlike in the inactive region—is non-zero.

Let $\rho$ be the mass density (per unit length) of the ideal tube and $g$ be the gravitational acceleration constant.
We prescribe $\mathcal{V} = \rho g \bE_1\cdot\br$ to be the gravitational potential.
The director components of the force and moment balance can then be written using \eqref{eq:director_orthonormality} and \eqref{eq:planar_representations} as
\begin{subequations}\label{eq:catenary_force_moment_balance_components}
\begin{align}
n_1' &= -u_2 n_3 + \rho g\cos\theta\, ,&\,\,
n'_3 &= u_2 n_1 + \rho g \sin\theta\, ,&\,\, 
0 &= -n_1\, ,&\,\, &\forall s\in(0,L)\setminus(s_1,s_2)\, ,\label{eq:catenary_inactive_force_moment_balance_components}
\\
n'_1 &= -\hat\kappa n_3 + \rho g \cos\theta\,,&\,\,
n'_3 &= \hat\kappa n_1 + \rho g \sin\theta\,,&\,\,
m'_2 &=-n_1\,,&\,\, &\forall s\in(s_1,s_2)\, .\label{eq:catenary_active_force_moment_balance_components}
\end{align}
\end{subequations}
Both sets of equations can be integrated analytically in conjunction with \eqref{eq:2D_kinematics}.
The solution in the inactive region is given by
\begin{equation}\label{eq:catenary_unactivated_integrated}
\begin{rcases}
\begin{array}{rl}
&r_1= \sqrt{c_3^2 + \left(c_1+s\right)^2} + o_1\, ,\\[2ex]
&r_3= c_3\arcsinh\left(\dfrac{c_1 + s}{c_3}\right) + o_3\, ,\\[2ex]
&\theta=\arctan\left(\dfrac{c_1 + s}{c_3}\right)\, , \\[2ex]
&n_1=0\, ,\\[2ex]
&n_3=\dfrac{c_3\rho g}{\cos\theta}\, ,\\[2ex]
&m_2=0\, .
\end{array}
\end{rcases}\forall s\in(0,L)\setminus(s_1,s_2)\, .
\end{equation}
Similarly, in the active region, the solution reads
\begin{equation}\label{eq:catenary_activated_integrated}
\begin{rcases}
\begin{array}{rl}
r_1 &= -\dfrac{\cos\theta}{\hat\kappa} + o_1\, ,\\[2ex]
r_3 &= \dfrac{\sin\theta}{\hat\kappa}+o_3\, ,\\[2ex]
\theta &=\hat\kappa s + \theta_0\, ,\\[2ex]
n_1 &=(c_1+\rho g s)\cos\theta - c_3 \sin\theta\, ,\\[2ex]
n_3  &= (c_1+\rho g s)\sin\theta - c_3 \cos\theta\, ,\\[2ex]
m_2 &= -\dfrac{(c_1+\rho g s)\sin\theta}{\hat\kappa} - \dfrac{(c_3\hat\kappa + \rho g)\cos\theta}{\hat\kappa^2} + c_0\, .
\end{array}
\end{rcases}\forall s\in(s_1,s_2)\, ,
\end{equation}
where $o_1,o_3,\theta_0,c_1,c_3$, and $c_0$ are constants of integration.
\begin{figure}
	 \centering
		\captionsetup[subfigure]{justification=centering}
	\begin{subfigure}[t]{0.25\textwidth}
		\includegraphics[trim={5.5cm 0.0cm 6.1cm 0.5cm}, clip,width=\linewidth]      {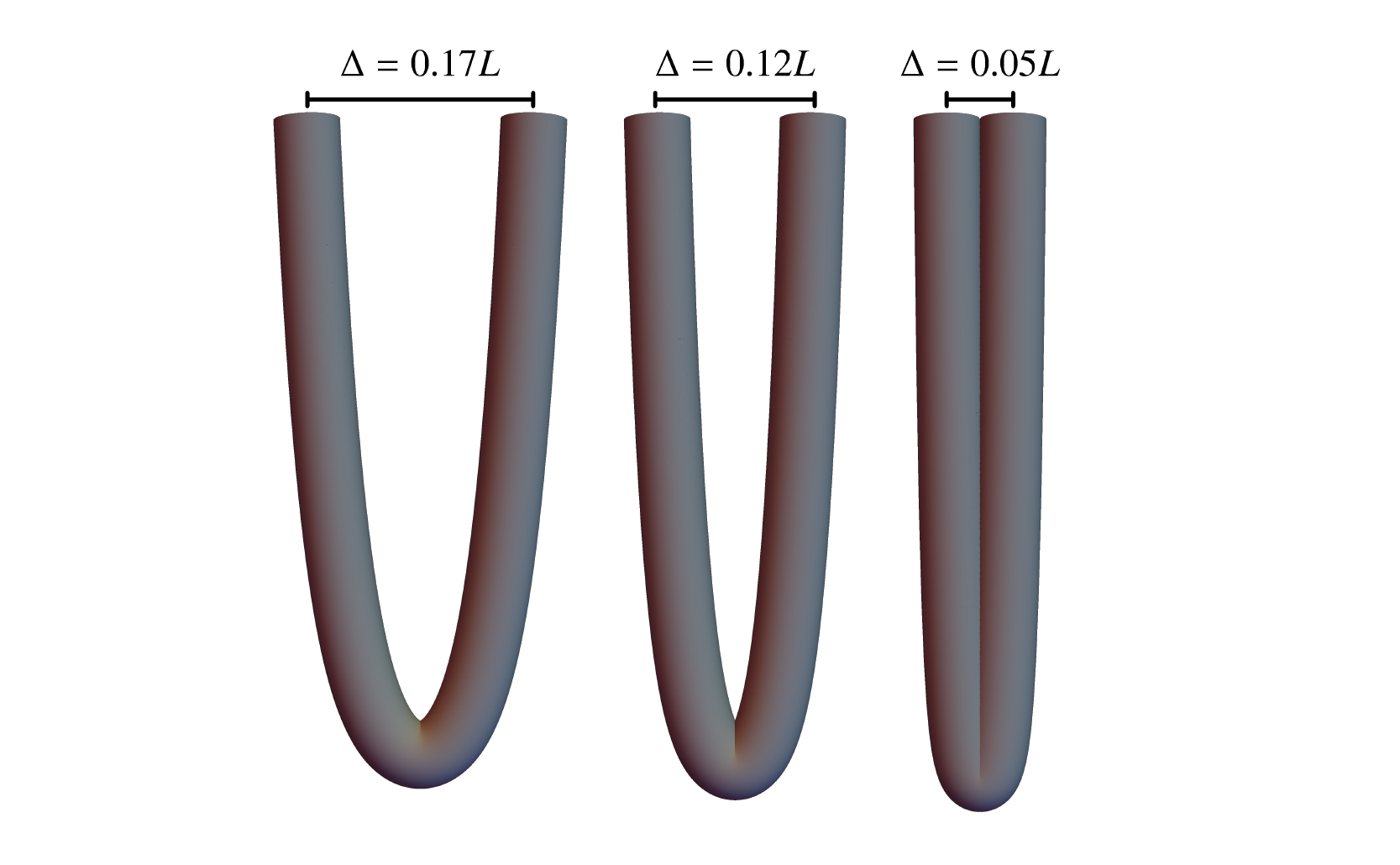}
		\caption{}
		\label{fig:idealtube_config1}
        \end{subfigure}%
	\hfill
	\begin{subfigure}[t]{0.37\textwidth}
		\includegraphics[width=0.94\linewidth]{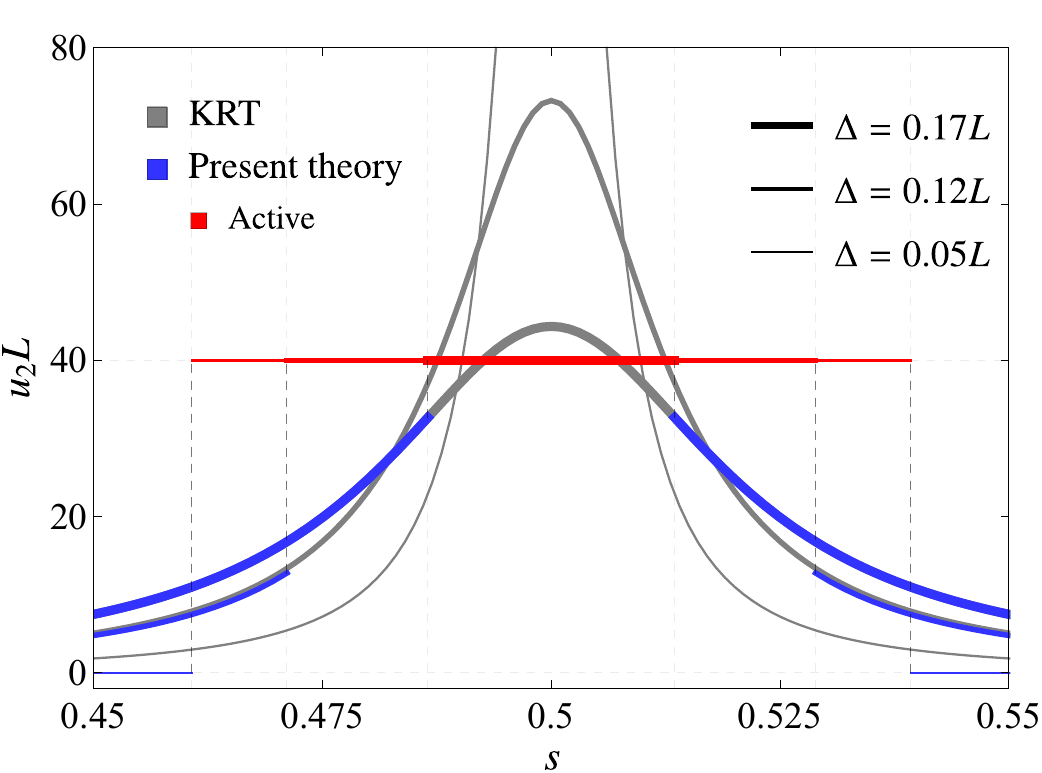}
		\caption{}
		\label{fig:idealtube_u2}
	\end{subfigure}%
    \hfill
	\begin{subfigure}[t]{0.37\textwidth}
		\includegraphics[width=0.95\linewidth]{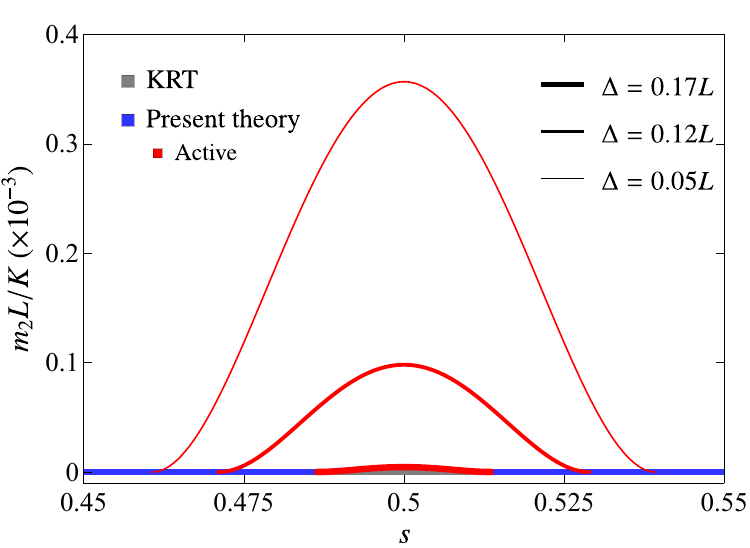}
		\caption{}
		\label{fig:idealtube_m2}
	\end{subfigure}
	\vspace{1em} 
	\begin{subfigure}[t]{0.25\textwidth}
			\includegraphics[trim={5.5cm 0.0cm 6.1cm 0.5cm}, clip,width=\linewidth]      {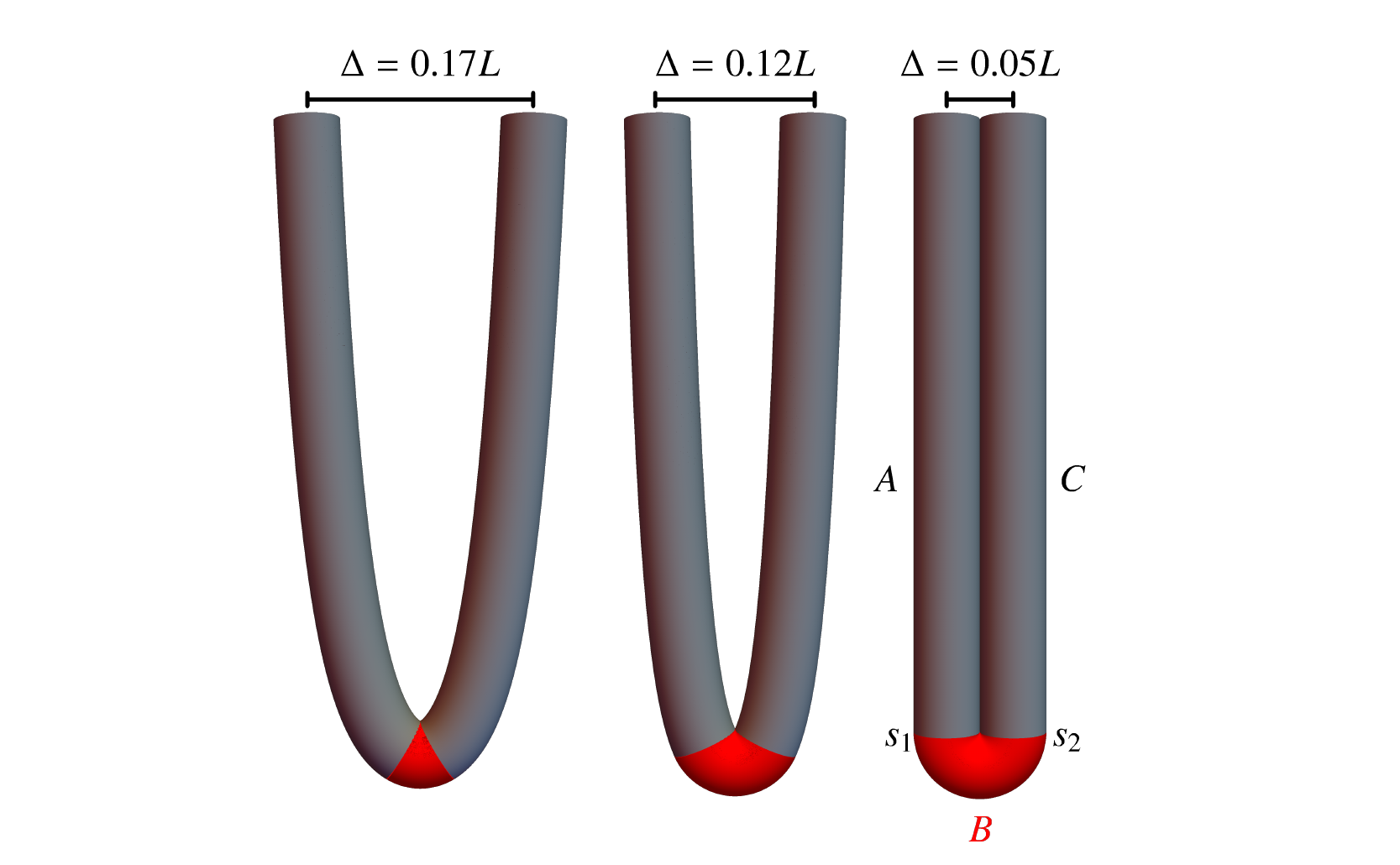}
		\caption{}
		\label{fig:idealtube_config2}
	\end{subfigure}%
	\hfill
	\begin{subfigure}[t]{0.37\textwidth}
		\includegraphics[width=\linewidth]{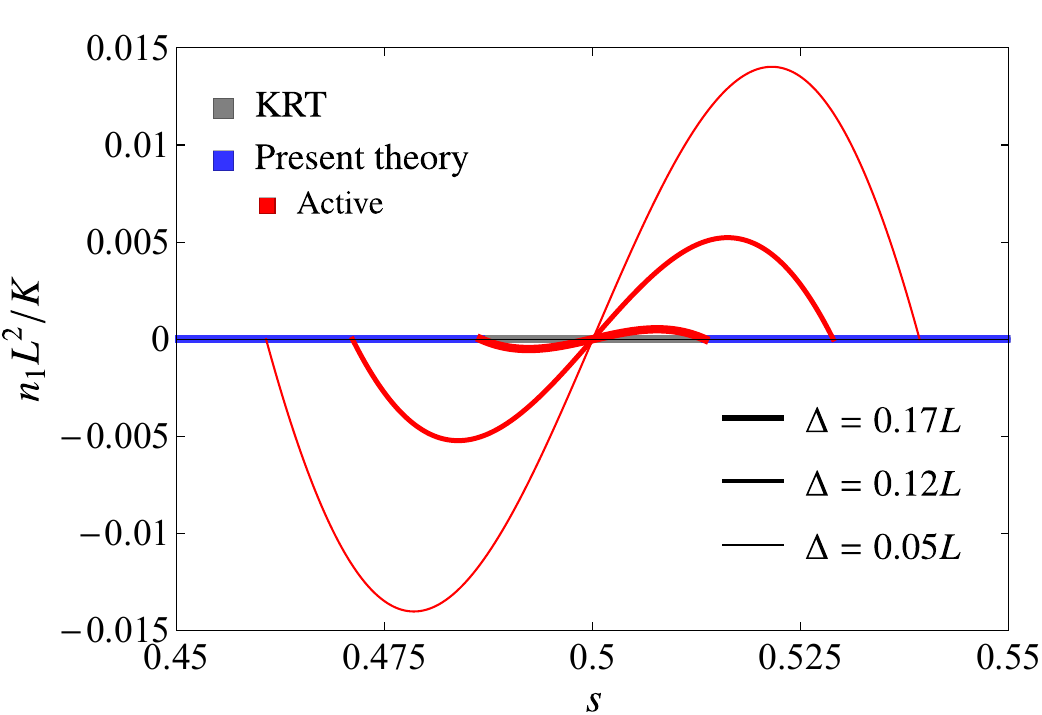}
		\caption{}
		\label{fig:idealtube_n1}
	\end{subfigure}%
	\hfill
	\begin{subfigure}[t]{0.37\textwidth}
		\includegraphics[width=0.98\linewidth]{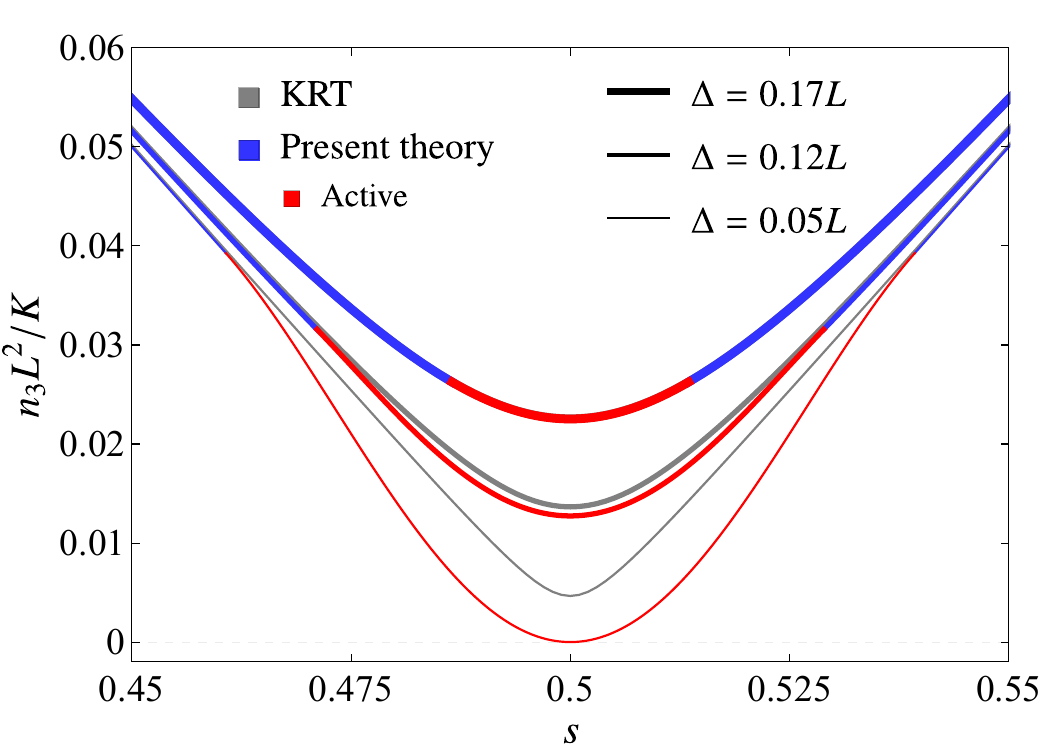}
		\caption{}
		\label{fig:idealtube_n3}
	\end{subfigure}
	\caption{The equilibrium configurations and mechanical fields for a fully flexible tube of thickness $t=0.025L$ hanging under self-weight. Panels \ref{fig:idealtube_config1} and \ref{fig:idealtube_config2} display the configurations of a standard fully flexible tube and a locally impenetrable ideal tube, respectively. The variations of the normalized curvature $u_2L$, bending moment $m_2L/K$, shear force $n_1L^2/K$, and axial force $n_3L^2/K$ along the length of the tube are shown in panels \ref{fig:idealtube_u2}, \ref{fig:idealtube_m2}, \ref{fig:idealtube_n1}, and \ref{fig:idealtube_n3}. Differences between the two solutions are negligible away from the active region and are omitted for clarity.}
\label{fig:idealtube}
\end{figure}

In order to compute the full configuration, we divide the entire domain into three parts: (i) Region A, where $s\in(0,s_1)$, (ii) Region B, where $s\in(s_1,s_2)$, and (iii) Region C, where $s\in(s_2,L)$ (see Fig. \ref{fig:idealtube_config2}).
Regions $A$ and $C$ are inactive, whereas Region $B$ is active.
Therefore, equations \eqref{eq:catenary_unactivated_integrated} govern regions $A$ and $C$, while region $B$ is governed by \eqref{eq:catenary_activated_integrated}.
The entire problem then reduces to computing the constants of integration along with $s_1$ and $s_2$, which locate the boundaries of the active region $B$. 
The complete state vector of unknowns is given by $\{o^A_1,o^A_3,c^A_1,c^A_3,o^B_1,o^B_3,\theta^B_0,c^B_0,c^B_1,c^B_3,o^C_1,o^C_3,c^C_1,c^C_3,s_1,s_2\}$, comprising $16$ unknowns, where the superscripts identify the region to which each constant corresponds.
We therefore require a total of $16$ boundary conditions.
Four of these are determined by fixing the locations of the two terminal ends of the tube:
\begin{align}\label{eq:catenary_boundary_conditions}
    r^A_1(0) = 0\, ,\quad
    r^A_3(0) = 0\, ,\quad
    r^C_1(L) = 0\, ,\quad
    r^C_3(L) =\Delta\, .
\end{align}
The remaining $12$ conditions come from the jump conditions valid at the interfaces $s_1$ and $s_2$.
These comprise the continuity of position and tangent vector \eqref{eq:position_tangent_jump_conditions}, alongside the internal force and moment balances \eqref{eq:force_moment_jump_conditions}. 
Projecting \eqref{eq:position_tangent_jump_conditions}$_1$ along $\{\bE_1,\bE_3\}$, and \eqref{eq:position_tangent_jump_conditions}$_2$ and \eqref{eq:force_moment_jump_conditions} along $\{\bd_1,\bd_3\}$, delivers the following matching relations:
\begin{subequations}\label{eq:catenary_connecting_conditions}
    \begin{align}
        \left(r_1^B - r_1^A\right)\big|_{\{s_1,s_2\}}&=0,&\quad \left(r_3^B - r_3^A\right)\big|_{\{s_1,s_2\}}&=0,&\quad \left(\theta^B - \theta^A\right)\big|_{\{s_1,s_2\}}&=0,
        \\
        \left(n_1^B - n_1^A\right)\big|_{\{s_1,s_2\}}&=0,&\quad \left(n_3^B - n_3^A\right)\big|_{\{s_1,s_2\}}&=0 \, ,&\quad \left(m_2^B - m_2^A\right)\big|_{\{s_1,s_2\}}&=0\, .
    \end{align}
\end{subequations}
The jump condition on the Hamiltonian in this case is automatically satisfied by the continuity of the moment, tangent, and internal force; it therefore offers no additional information.
Finally, the Lagrange multiplier $\Lambda$ can be evaluated via $\Lambda=m_2$, as implied by \eqref{eq:piecewise_constitutive_relation_string}.

We compare the solutions computed using the present theory with the standard catenary solution for an ideal tube of radius $t=0.025L$ hanging under self-weight, as shown in Fig. \ref{fig:idealtube}.
The curvature at the lowest point hits its peak value $\hat\kappa L = 1/0.025 = 40$ at $\Delta = 0.184L$.
As $\Delta$ is reduced further, local penetration in the standard solution becomes apparent in Fig. \ref{fig:idealtube_config1}.
Configurations computed using the present theory are presented in Fig. \ref{fig:idealtube_config2}.
Apart from the differences in the shapes of the two sets of configurations, several qualitative distinctions are worth highlighting.

The curvature $u_2$ of the locally impenetrable flexible tube suffers a jump at the boundary of the active region (colored red), as can be seen in Fig. \ref{fig:idealtube_u2}, while in standard catenary solutions, the curvature remains continuous throughout. 
The existence of jumps in curvature is consistent with the regularity results proved in \citet{gonzalez2002} for self-contacting tubes.
The locally impenetrable flexible tube is also able to support internal moments $m_2$ and shear forces $n_1$ within the active region, as demonstrated in Figs. \ref{fig:idealtube_m2} and \ref{fig:idealtube_n1}.
Finally, we note that the tension $n_3$ in the active region is lower when compared to the tensions predicted by the standard catenary solutions.
All of these differences are direct consequences of the change in the constitutive description induced by the local impenetrability constraint across the active and inactive regions of the flexible tube.

Interestingly, one fundamental property of the standard catenary solution survives the addition of the impenetrability constraint:
the shapes predicted by both the standard solution and the present theory remain independent of the strength of the gravitational field.
This can be verified by non-dimensionalizing equations \eqref{eq:catenary_force_moment_balance_components} using a length scale $L$ and a force scale $\rho g L$, which leads to two non-dimensional sets of equations that contain no free parameters.

One might expect that the addition of bending elasticity to a flexible tube hanging under self-weight would regularize curvatures enough to prevent local penetration entirely.
We show in the next section that this holds true only if the bending elasticity of the rod exceeds a certain threshold.

\section{An elastic tube hanging under self-weight}\label{sec:elastica_under_gravity}

Let us consider an \emph{elastic tube} of radius $t$, length $L$, and bending modulus $K$ hanging under self-weight with its two ends pinned at a horizontal distance $\Delta$.
As in the case of a fully flexible tube, the peak curvature in the elastic tube would occur at the lowest point on the centerline of the tube. 
The magnitude of the peak curvature would be a function of the bending modulus $K$.
The lower the bending modulus, the higher the peak curvature.
Our interest here is in finding the threshold of the bending modulus under which the peak curvature reaches the upper bound $\hat\kappa$ for some value of $\Delta$, and computing configurations when $\Delta$ is reduced further.

The kinematics and planar representations of various fields stated in \eqref{eq:planar_representations} remain applicable in this case.
\begin{figure}[t]
	\centering
	\captionsetup[subfigure]{justification=centering}
	\begin{subfigure}[t]{0.25\textwidth}
		\includegraphics[trim={7.0cm 0.0cm 6.5cm 0.5cm}, clip,width=0.85\linewidth]      {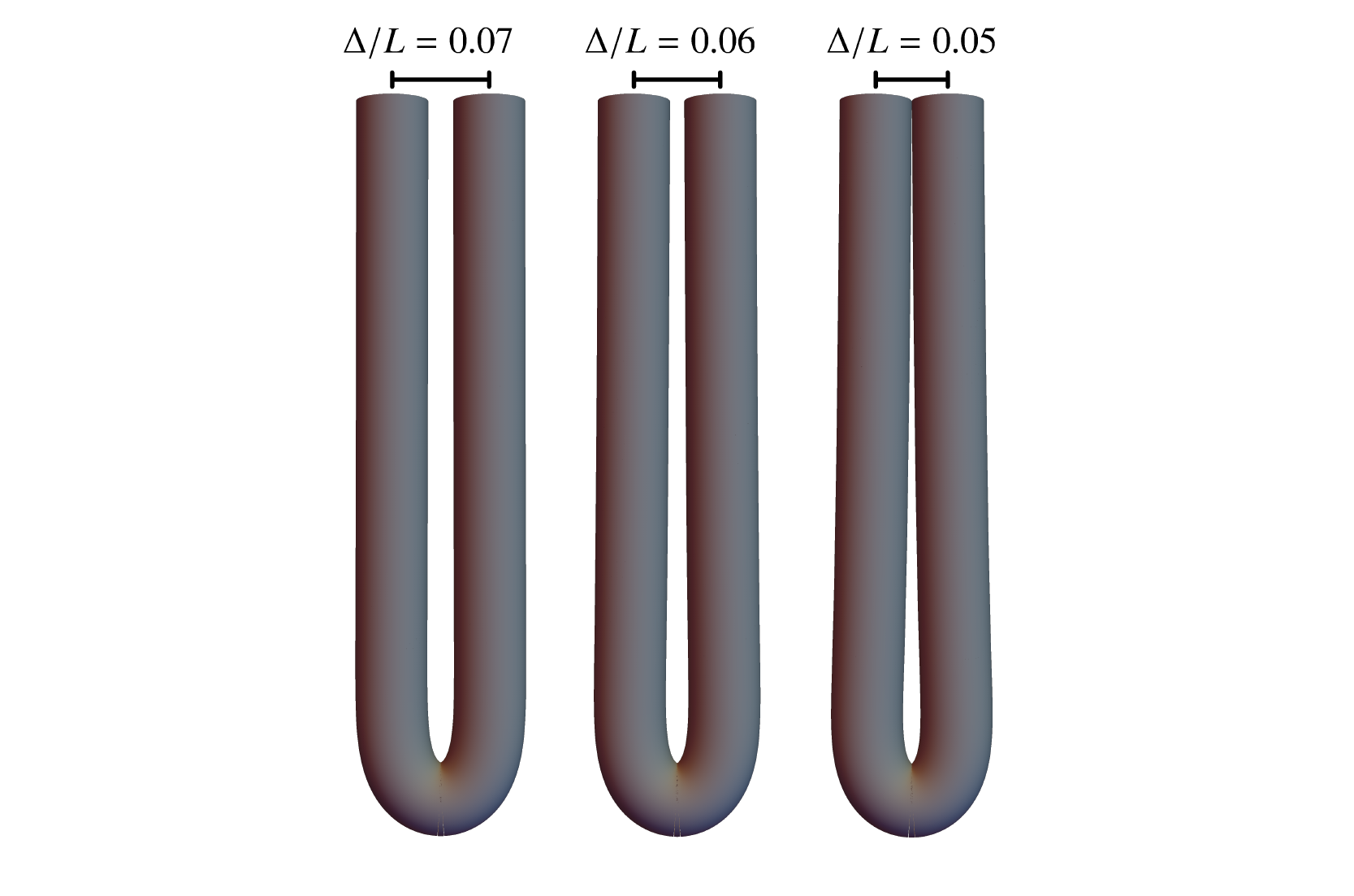}
		\caption{}
		\label{fig:elastica_config1}
	\end{subfigure}%
	\hfill
	\begin{subfigure}[t]{0.37\textwidth}
		\includegraphics[width=0.95\linewidth]{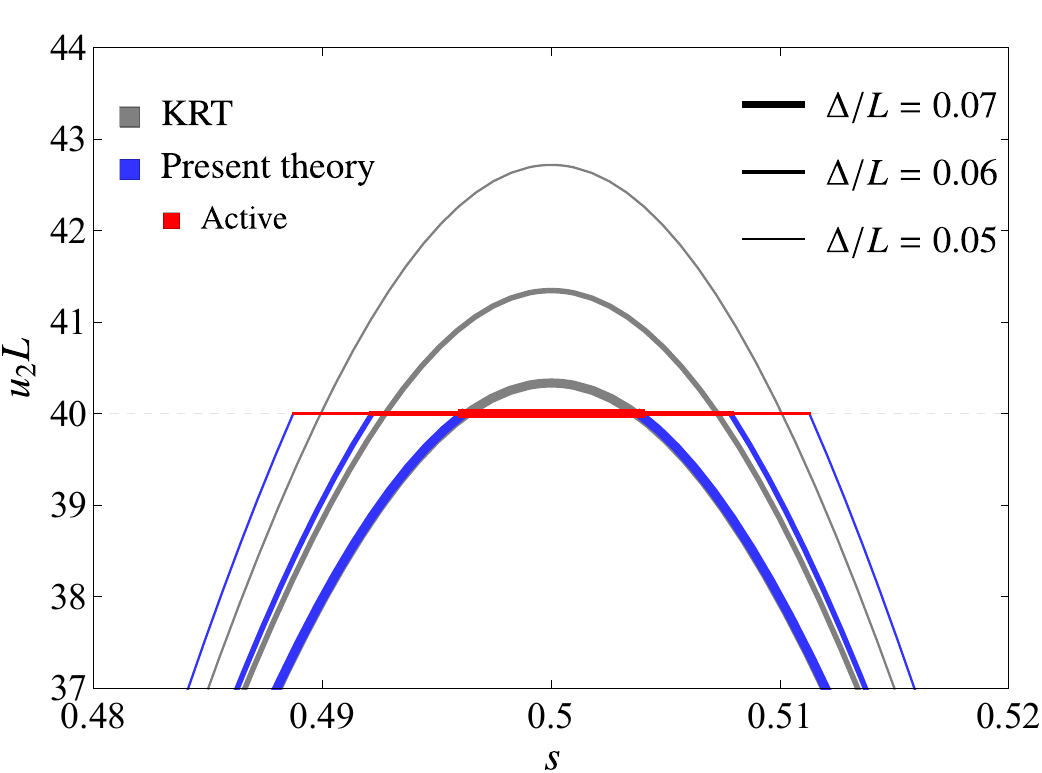}
		\caption{}
		\label{fig:elastica_u2}
	\end{subfigure}%
	\hfill
	\begin{subfigure}[t]{0.37\textwidth}
		\includegraphics[width=\linewidth]{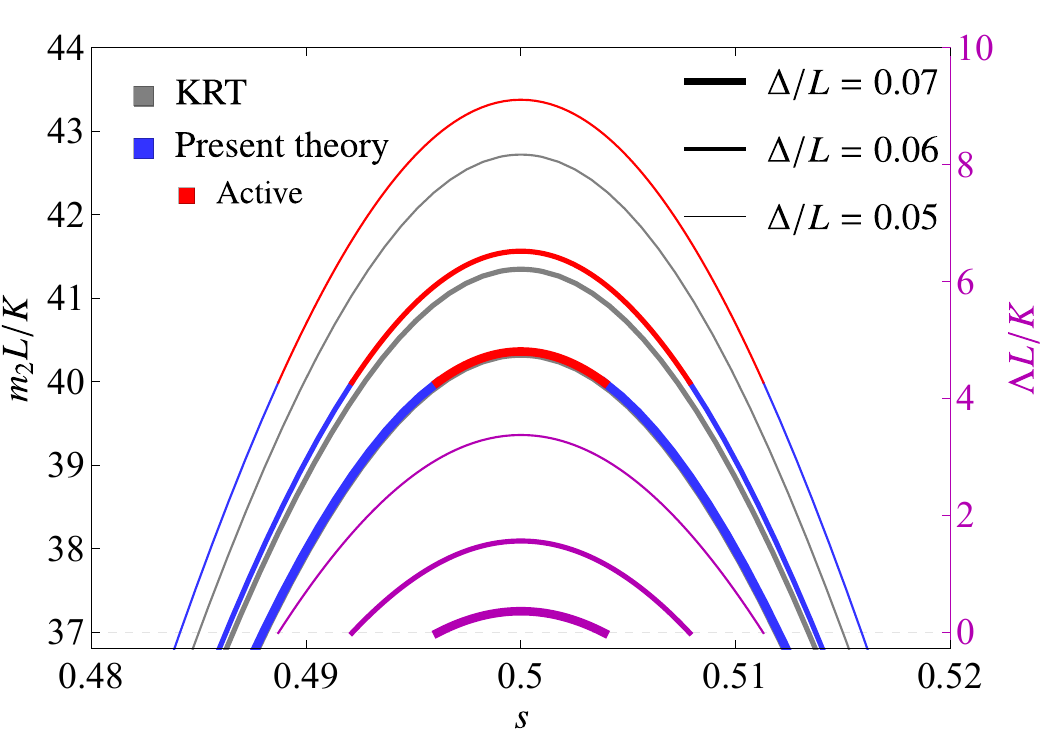}
		\caption{}
		\label{fig:elastica_m2}
	\end{subfigure}
	
	\vspace{1em} 
	
	\begin{subfigure}[t]{0.25\textwidth}
		\includegraphics[trim={7.0cm 0.0cm 6.5cm 0.5cm}, clip,width=0.85\linewidth]      {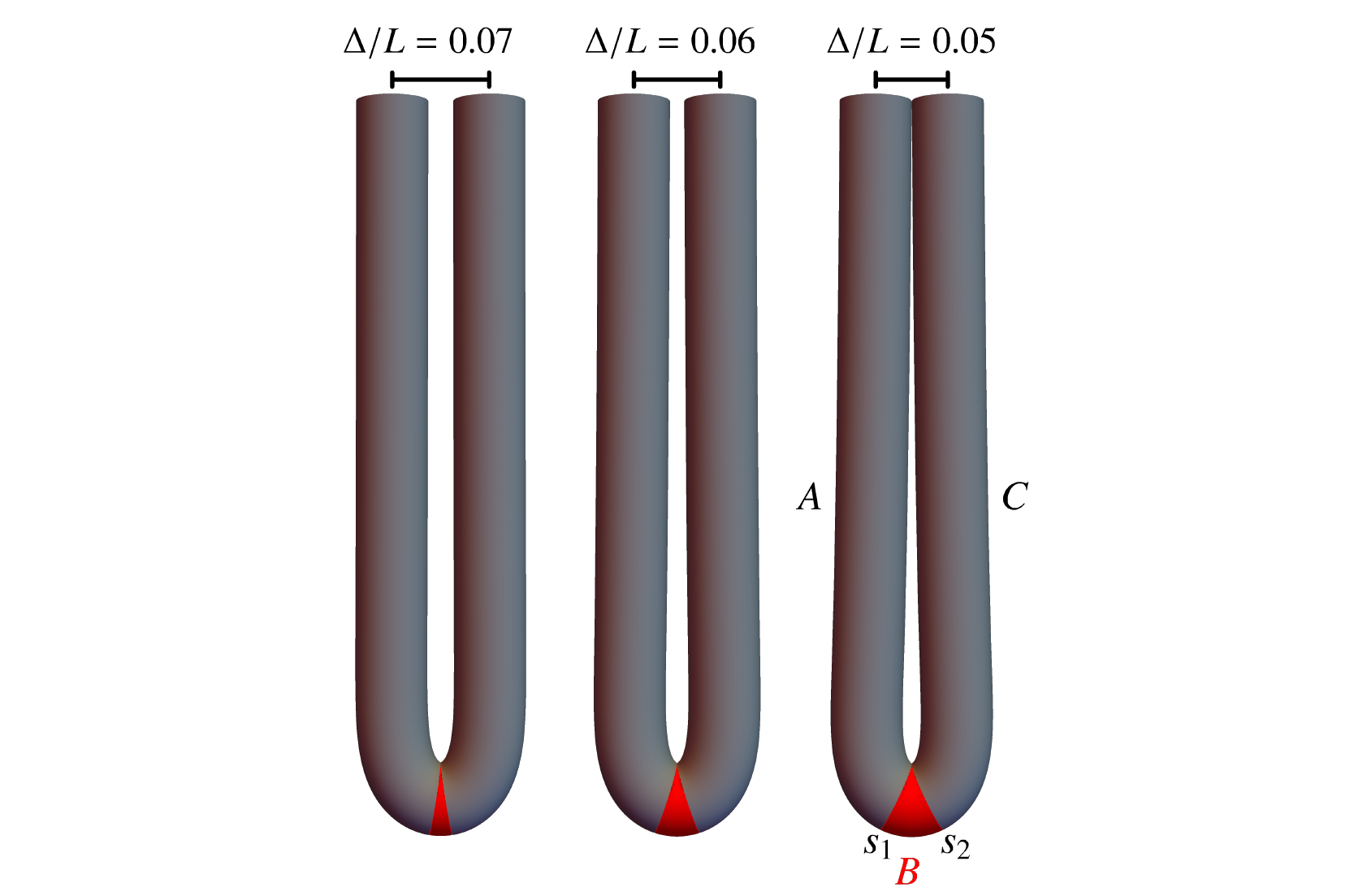}
		\caption{}
		\label{fig:elastica_config2}
	\end{subfigure}%
	\hfill
	\begin{subfigure}[t]{0.37\textwidth}
		\includegraphics[width=\linewidth]{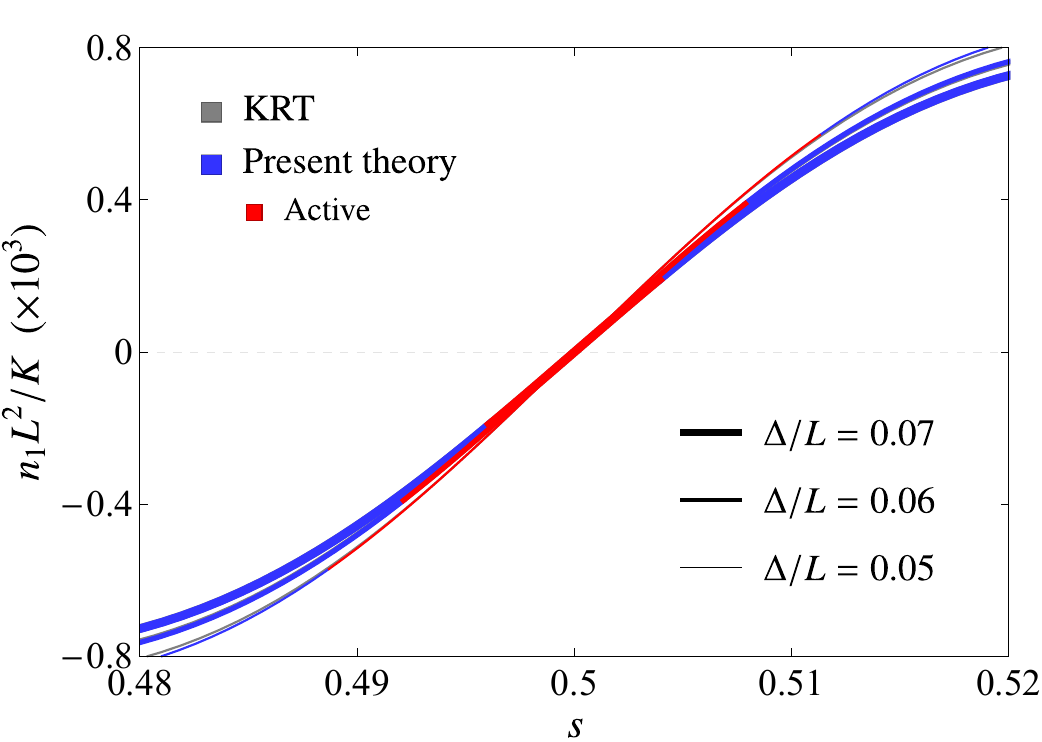}
		\caption{}
		\label{fig:elastica_n1}
	\end{subfigure}%
	\hfill
	\begin{subfigure}[t]{0.37\textwidth}
		\includegraphics[width=0.95\linewidth]{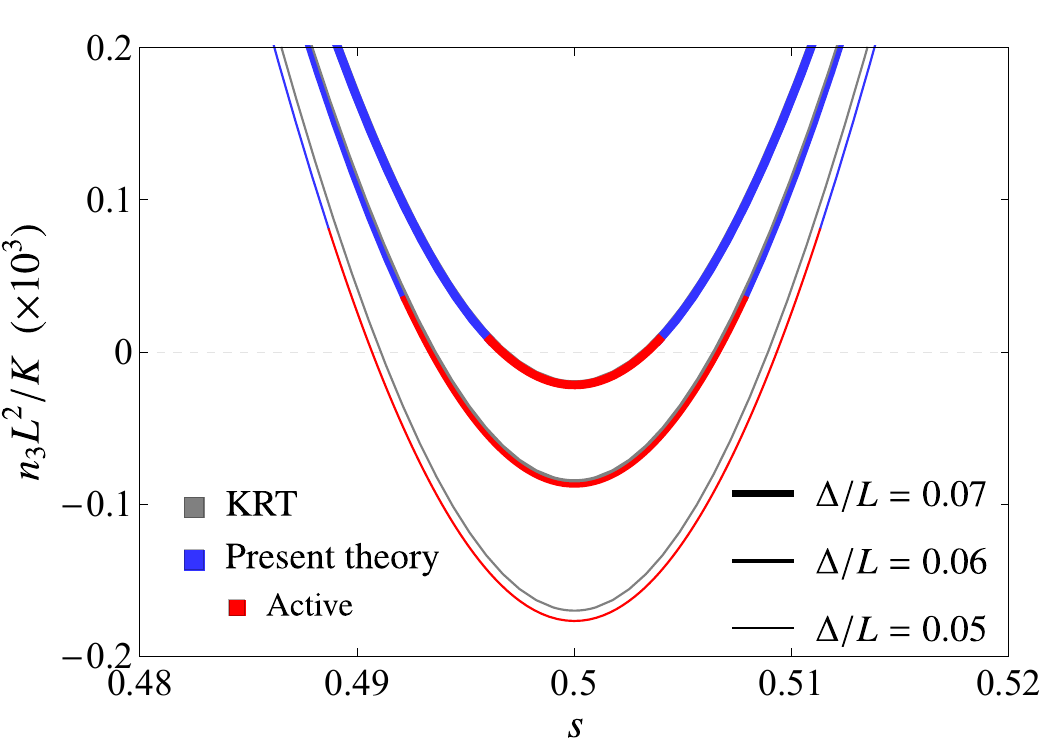}
		\caption{}
		\label{fig:elastica_n3}
	\end{subfigure}
	
	\caption{Solutions of a pinned elastica of thickness $t/L=0.025$ hanging under self-weight with $\alpha=4.86 \times 10^4$ and varying values of $\Delta/L$. The configurations obtained from standard KRT and the present theory are shown in panels \ref{fig:elastica_config1} and \ref{fig:elastica_config2}. Panels \ref{fig:elastica_u2} and \ref{fig:elastica_m2} show the respective variations of normalized curvature $u_2L$, normalized moment $m_2L/K$, and the parameter $\Lambda L/K$ along the length of the tube close to the active region (in red). The variations of shear force $n_1L^2/K$ and axial force $n_3L^2/K$ are shown in panels \ref{fig:elastica_n1} and \ref{fig:elastica_n3}.}
	\label{fig:elastica}
\end{figure}
The constitutive relation for the elastic tube can be written using \eqref{eq:piecewise_constitutive_relation} and \eqref{eq:Kirchhoff_energy_density} with $u_1=u_3=0$
\begin{align}\label{eq:piecewise_constitutive_relation_elastica}
\bmo = 
    \begin{dcases}
        K u_2 \bd_2 \quad &\text{for} \quad s \in (0,L) \setminus (s_1,s_2) \\
        (K \hat\kappa + \Lambda) \bd_2 \quad &\text{for} \quad s \in (s_1,s_2)\, ,
    \end{dcases}
\end{align}
where $s \in (s_1,s_2)$ is identified with the active region in the tube.

The force and moment balance in the inactive and active regions can be written using \eqref{eq:force_moment_balance} with the gravitational potential $\mathcal{V} = \rho g \bE_1\cdot\br$.
Using \eqref{eq:piecewise_constitutive_relation_elastica} to eliminate $u_2$ in favor of $m_2$, the full set of ODEs applicable in the inactive region are given by
\begin{equation}\label{eq:force_moment_balance_scalar_inactive}
    \begin{rcases}
        \begin{aligned}
           r'_1&=\sin\theta\, ,&\qquad r'_3&=\cos\theta\, ,&\qquad\theta'&=\frac{m_2}{K}\, ,\\
           n'_1&=-\frac{m_2}{K} n_3 + \rho g\cos\theta\, ,& n'_3&=\frac{m_2}{K} n_1 + \rho g \sin\theta\, ,&
           m'_2&=-n_1\, ,
       \end{aligned}
    \end{rcases} \forall s \in (0,L) \setminus (s_1,s_2)\, .
\end{equation}
In the activated region, the curvature $u_2=\hat\kappa$ from \eqref{eq:active_region}, and therefore the governing set of ODEs are given by,
\begin{equation}\label{eq:force_moment_balance_scalar_active}
    \begin{rcases}
        \begin{aligned}
           r'_1&=\sin\theta\, ,&\qquad r'_3&=\cos\theta\, ,&\qquad\theta'&=\hat\kappa\, ,\\
           n'_1&=-\hat\kappa n_3 + \rho g\cos\theta\, ,& n'_3&=\hat\kappa n_1 + \rho g \sin\theta\, ,&
           m'_2&=-n_1\, ,
       \end{aligned}
    \end{rcases} \forall s \in (s_1,s_2)\, .
\end{equation}
Similar to Section \ref{sec:ideal_tube}, we divide the length of the tube into three regions, namely region A, B, and C, where equations \eqref{eq:force_moment_balance_scalar_inactive} are valid in regions $A$ and $C$ and \eqref{eq:force_moment_balance_scalar_active} are valid in region $B$.
Using these equations, the configuration can be solved by constructing a boundary value problem in the following unknown vector with 18 unknown functions and 2 unknown parameters: $\{r_1^A,r_3^A,\theta^A,n_1^A,n_3^A,m_2^A,r_1^B,r_3^B,\theta^B,n_1^B,n_3^B,m_2^B,r_1^C,r_3^C,\theta^C,n_1^C,n_3^C,m_2^C,s_1,s_2\}$.
Therefore, we require 20 boundary conditions to pose the full boundary value problem.

Six boundary conditions are prescribed as follows at the two terminal ends of the rod,
\begin{subequations}\label{eq:elastica_boundary_conditions}
\begin{align}
    r^A_1(0) &= 0\, ,\quad & r^A_3(0) &= 0\, ,\quad & m_2^A(0) &= 0\, , \\
    r^C_1(L) &= 0\, ,\quad & r^C_3(L) &= \Delta\, ,\quad & m_2^C(L) &= 0\, .
\end{align}
\end{subequations}
The remaining conditions arise from imposing the continuity of the position and the tangent \eqref{eq:position_tangent_jump_conditions}, the internal force and internal moment \eqref{eq:force_moment_jump_conditions}, along with the continuity of the Hamiltonian \eqref{eq:Hamiltonian_continuity} at the boundaries $\{s_1,s_2\}$ of the active region,
\begin{subequations}\label{eq:elastica_connecting_conditions}
    \begin{align} 
        \jump{r_1}\big|_{\{s_1,s_2\}} &= 0\, ,&\quad \jump{r_3}\big|_{\{s_1,s_2\}} &= 0\, ,&\quad \jump{\theta}\big|_{\{s_1,s_2\}} &= 0\, ,\\
        \jump{n_1}\big|_{\{s_1,s_2\}} &= 0\, ,&\quad \jump{n_3}\big|_{\{s_1,s_2\}} &= 0\, ,&\quad \jump{m_2}\big|_{\{s_1,s_2\}} &= 0\, ,\\
        \jump{H}\big|_{\{s_1,s_2\}} &= 0\, .
    \end{align}
\end{subequations}
We choose $L$ to be the length scale and $K/L^2$ to be the force scale to non-dimensionalize equations \eqref{eq:force_moment_balance_scalar_inactive} and \eqref{eq:force_moment_balance_scalar_active}, and state the full boundary value problem in Appendix \ref{app:bvp_elastic_tube}.
The relevant parameter that appears out of this exercise is $\alpha = \rho g L^3/K$.

The BVP is solved using a parameter continuation scheme implemented in \texttt{AUTO-07P} \cite{doedel1998}.
We present an example of an elastic tube of radius $t=0.025L$ pinned at its terminal ends.
The curvature at the lowest point hits its peak $\hat\kappa L = 1/0.025 = 40$ at $\Delta = 0.0697L$ for a value of $\alpha = 4.86 \times 10^4$, which corresponds to a very soft material\footnote{As an example, a tube made of polydimethylsiloxane (PDMS) with $E=0.57$~MPa, $\rho=980~\text{kg/m}^3$, $t=2.5$~mm, and $L=1.657$~m would result in $\alpha=4.86 \times 10^4$.} \cite{wang2014,brounstein2021}. 
Reducing $\Delta$ further leads to the nucleation and propagation of an active region of constant curvature.
The configurations without the impenetrability constraint enforced are shown in Fig.~\ref{fig:elastica}a, whereas configurations obtained using the present theory are shown in Fig.~\ref{fig:elastica}d.
The difference in the shape of the elastic tube in and around the active region is shown in the curvature plot in Fig.~\ref{fig:elastica_u2}.
Although not visible to the naked eye, there is a tiny jump in the curvature across the boundaries of the active region $s_1$ and $s_2$ in Fig.~\ref{fig:elastica_u2}, even though the internal moment is continuous.
The continuity of the internal moment can be written as $\jump{Ku_2 + \Lambda} = 0$ by use of \eqref{eq:moment_jump_condition} and \eqref{eq:piecewise_constitutive_relation_elastica}, where a jump in $\Lambda$ must compensate for the jump in $u_2$.
As in the example of a fully flexible tube, the existence of jumps in the curvature of the centerline here is consistent with the regularity results proved by \citet{gonzalez2002} for self-contacting tubes.
The differences in the magnitude of $m_2, n_1,$ and $n_3$ are very small when compared to solutions of the standard KRT, as can be seen in Figs.~\ref{fig:elastica_m2}, \ref{fig:elastica_n1}, and~\ref{fig:elastica_n3}.
This example demonstrates that for tubes made of extremely soft materials, bending elasticity alone may not be sufficient to compute accurate shapes.

Finally, in the next section, we demonstrate a 3D example of our theory comprising a highly twisted elastic rod and show how accounting for local impenetrability results in significant differences in the deformed configuration as compared to the standard Kirchhoff rod theory solutions.

\section{3D twisted solutions}\label{sec:3D_sol}

We consider an elastic tube of length $L$ and radius $t$, with $K_1 = K_2 = K$.
The translational and rotational degrees of freedom at $s=0$ are constrained such that the directors coincide with $\bE_i$, and the position vector coincides with the origin.
At the outset, we apply a pure twisting moment along $\bE_3$ at $s=L$, with the tangent there constrained to coincide with $\bE_3$, and translations permitted only along $\bE_3$.
As the twisting moment is incrementally increased, an instability causes the rod to buckle out of plane.
We then constrain the only remaining translational degree of freedom at $s=L$ and continue to increase the twisting moment, leading to curvatures concentrating towards the middle of the rod.
The translational degree of freedom is constrained so as to prevent the formation of a plectoneme \cite{purohit2008plectoneme} and to prevent global contact from occurring before local contact occurs.
If the upper bound on the Frenet curvature \eqref{eq:kappa_inequality} is ignored, the configurations resulting from the standard Kirchhoff rod theory lead to local penetration.
We compute configurations using the present theory and highlight its differences from the predictions of the Kirchhoff rod theory.

We use Euler-parameters $q_i\equiv q_i(s)$, where $i\in\{1,2,3,4\}$, subject to the unit-norm constraint $q_1^2+q_2^2+q_3^2+q_4^2 = 1$, to parametrize 3D rotations of the director frame.
The Cartesian components of the director are represented as
\begin{align}\label{eq:3D_director_parametrization}
    \bd_1 = \begin{pmatrix}
             q_1^2 - q_2^2 - q_3^2 + q_4^2\\
             2(q_1q_2+q_3q_4)\\
             2(q_1q_3-q_2q_4)
            \end{pmatrix}\,, \qquad
    \bd_2 = \begin{pmatrix}
              2(q_1q_2-q_3q_4)\\
              -q_1^2+q_2^2-q_3^2+q_4^2\\
              2(q_2q_3+q_1q_4)
            \end{pmatrix}\, ,\qquad
    \bd_3 = \begin{pmatrix}
              2(q_1q_3+q_2q_4)\\
              2(q_2q_3-q_1q_4)\\
              -q_1^2-q_2^2+q_3^2+q_4^2
            \end{pmatrix}\, .
\end{align}
The equations for the evolution of the Euler-parameters w.r.t. $s$ are obtained from \eqref{eq:director_orthonormality} with \eqref{eq:3D_director_parametrization} and the unit-norm constraint
\begin{align}\label{eq:3D_euler_parameters_evolution}
    \begin{pmatrix}
        q'_1\\
        q'_2\\
        q'_3\\
        q'_4
    \end{pmatrix}
    =\frac{1}{2}\begin{pmatrix}
                 0 & u_3 & -u_2 & u_1\\
                 -u_3 & 0 & u_1 & u_2\\
                 u_2 & -u_1 & 0 & u_3\\
                 -u_1 & -u_2 & -u_3 & 0
                \end{pmatrix}
                \begin{pmatrix}
                    q_1\\
                    q_2\\
                    q_3\\
                    q_4
                \end{pmatrix}\, .
\end{align}
\begin{figure}[t]
	\centering
	\captionsetup[subfigure]{justification=centering}
	\begin{subfigure}[t]{0.6\textwidth}
		\includegraphics[trim={6.2cm 6cm 3.5cm 6cm}, clip,width=\linewidth]      {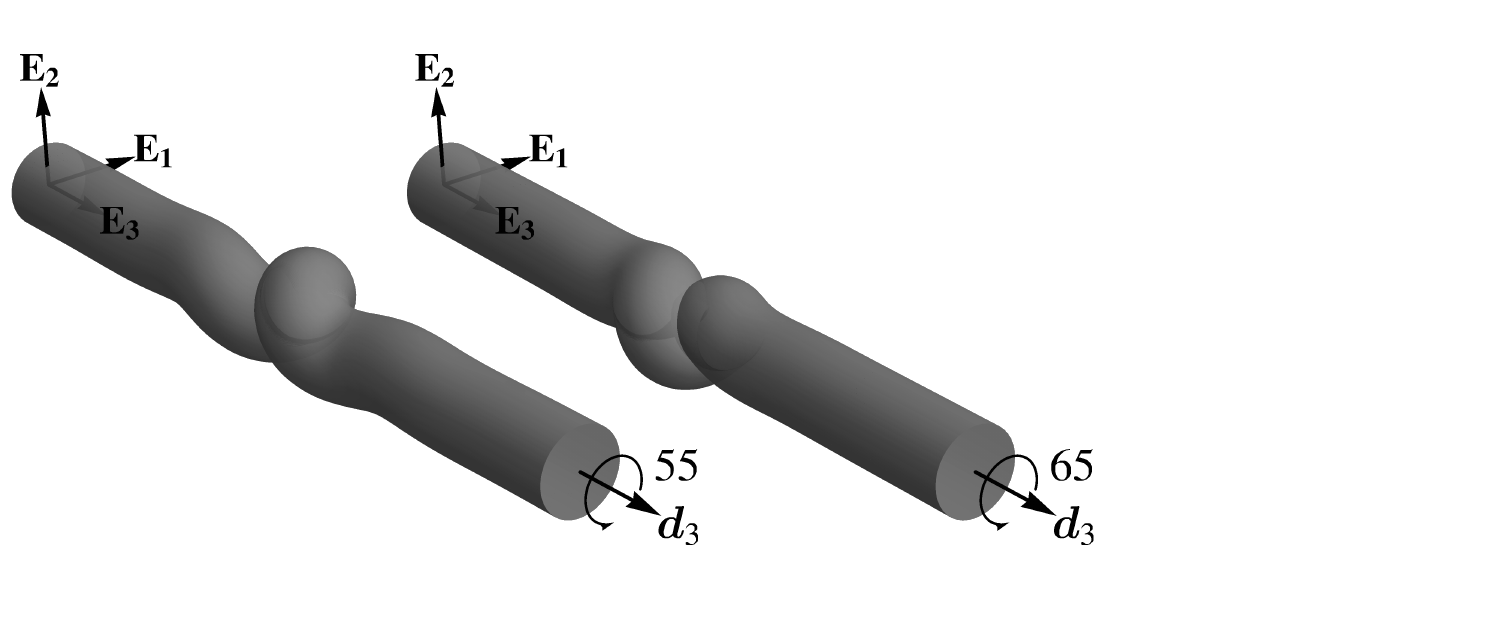}
		\caption{}
		\label{fig:rod_config1}
	\end{subfigure}%
	\begin{subfigure}[t]{0.37\textwidth}
		\includegraphics[width=0.95\linewidth]{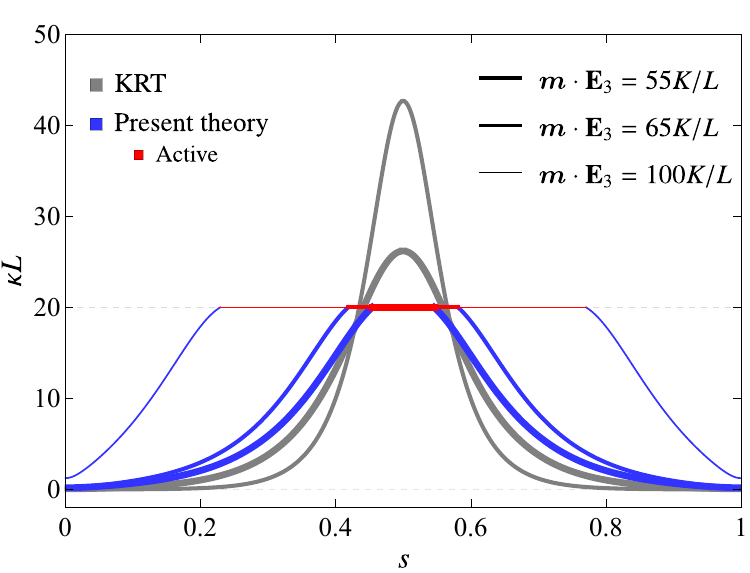}
		\caption{}
		\label{fig:rod_kappa}
	\end{subfigure}%
	\\	
	\begin{subfigure}[t]{0.6\textwidth}
		\includegraphics[trim={6.2cm 6.0cm 3.5cm 6.5cm}, clip,width=\linewidth] {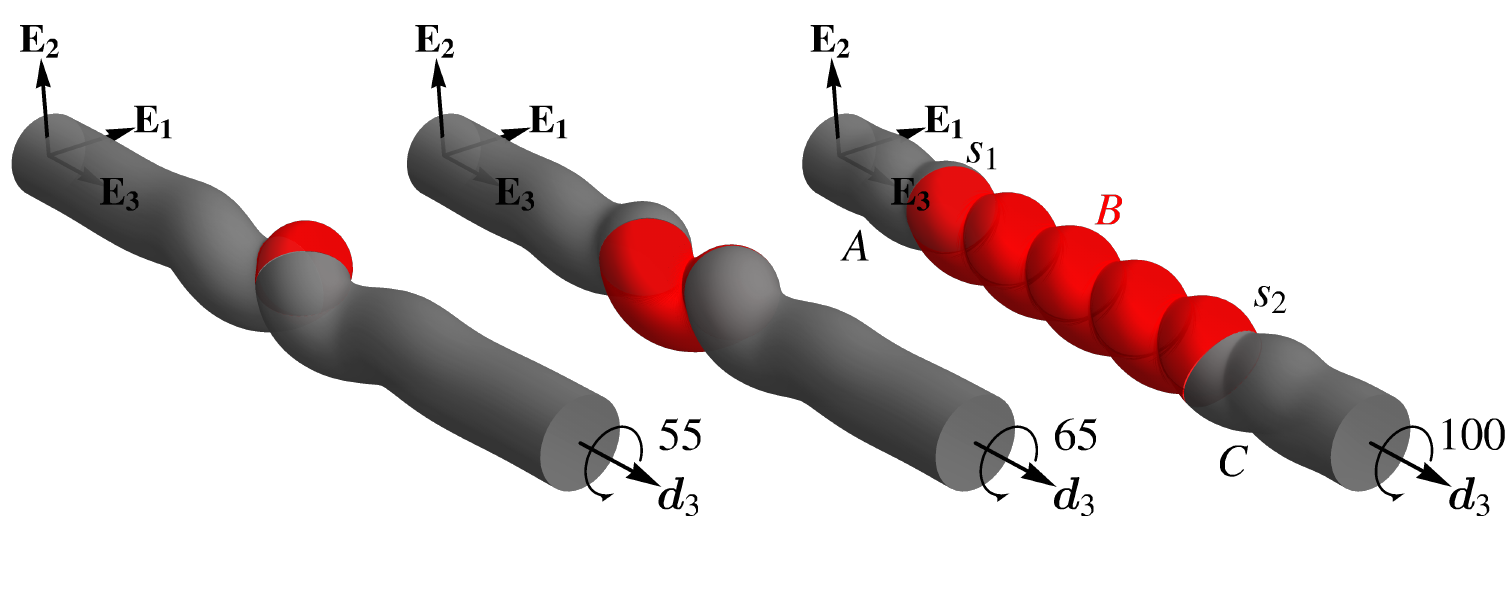}
		\caption{}
		\label{fig:rod_config2}
	\end{subfigure}%
	\begin{subfigure}[t]{0.37\textwidth}
		\includegraphics[width=0.95\linewidth]{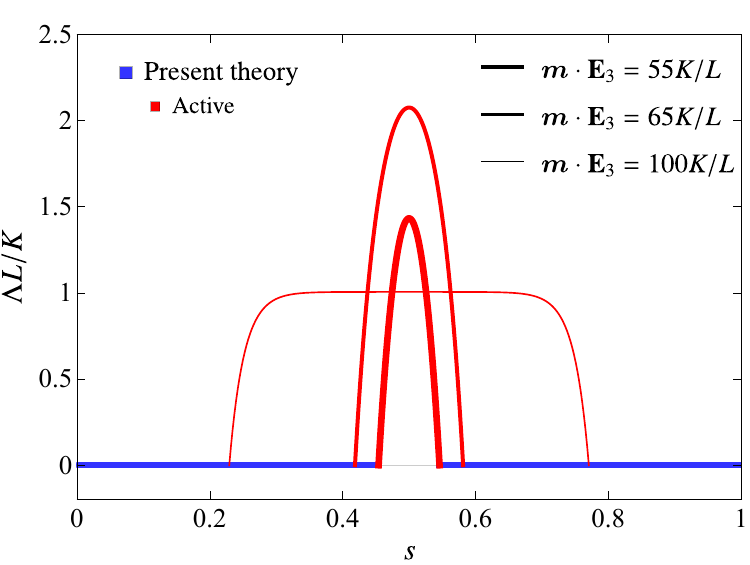}
		\caption{}
		\label{fig:rod_lambda}
	\end{subfigure}%
	\caption{Twisted configurations of a Kirchhoff rod and a locally impenetrable elastic tube with radius $t/L=0.05$ and $K_3/K=1$ are shown in panels \ref{fig:rod_config1} and \ref{fig:rod_config2}. The position and orientation of the rod are fixed at $s=0$. At $s=1$, the position is fixed, director $\bd_3$ is constrained to be along $\bE_3$, and a twisting moment $\bmo \cdot \bE_3$ is applied. Panels \ref{fig:rod_kappa} and \ref{fig:rod_lambda} show the variations along the length of the rod of normalized Frenet curvature $\kappa L$ and normalized parameter $\Lambda L/K$, respectively. The active region, in addition to having a constant Frenet curvature, also exhibits a constant value of geometric torsion under these loading conditions. Solutions obtained from the standard KRT do not converge beyond $m_3 \approx 66.11$.}
	\label{fig:rod_1}
\end{figure}
The components $r_i:=\br\cdot\bE_i$ of the centerline in the Cartesian basis are obtained using \eqref{eq:kirchhoff_kinematic} and \eqref{eq:3D_director_parametrization}$_3$,
\begin{align}\label{eq:3D_centerline}
    r'_1 = 2(q_1q_3 + q_2q_4)\, ,\qquad
    r'_2 = 2(q_2q_3 - q_1q_4)\, ,\qquad
    r'_3 = (-q_1^2-q_2^2+q_3^2+q_4^2)\, .\qquad
\end{align}
Let us consider an active region in the tube identified with the interval $(s_1,s_2)$.
Using \eqref{eq:piecewise_constitutive_relation} and \eqref{eq:Kirchhoff_energy_density}, the constitutive relation for such a rod can be written as

\begin{align}\label{eq:piecewise_constitutive_relation_twist}
\bmo = 
    \begin{dcases}
        Ku_1\bd_1+Ku_2\bd_2+K_3u_3\bd_3 \quad &\text{for} \quad s \in (0,L) \setminus (s_1,s_2) \\
        \left(K + \frac{\Lambda}{\hat\kappa}\right)u_1\bd_1+\left(K+\frac{\Lambda}{\hat\kappa}\right) u_2\bd_2+K_3u_3\bd_3 \quad &\text{for} \quad s \in (s_1,s_2)\, .
    \end{dcases}
\end{align}

Alternatively, the constitutive relation in the active region can also be written by expressing the bending strains in terms of the internal moments as in \eqref{eq:nonlinear_constitutive_relation}:
\begin{align}\label{eq:nonlinear_constitutive_relation_3D}
    u_1 = \frac{m_1\hat\kappa}{\left(m_1^2 + m_2^2\right)^{1/2}}\, ,\qquad u_2 = \frac{m_2\hat\kappa}{\left(m_1^2 + m_2^2\right)^{1/2}}\, ,\qquad u_3 = \frac{m_3}{K_3}\, , \qquad \forall s \in (s_1,s_2).
\end{align}

\begin{figure}[t]
	\centering
	\captionsetup[subfigure]{justification=centering}
    \begin{subfigure}{0.32\textwidth}
		\includegraphics[width=\linewidth]{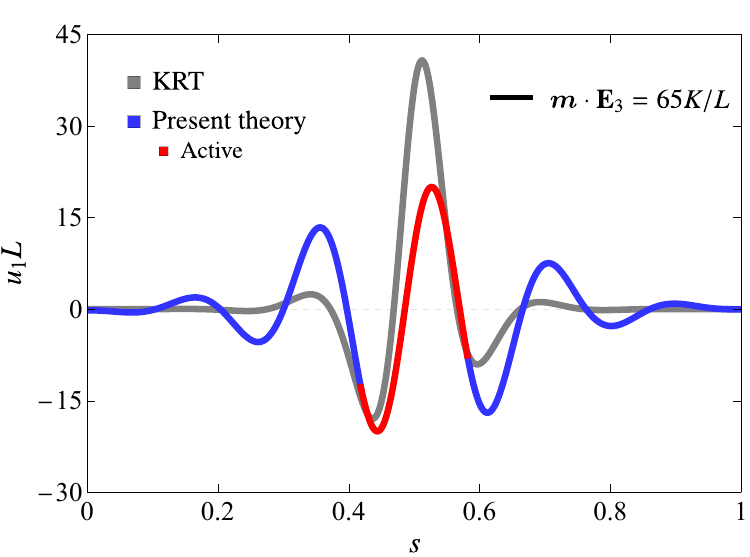}
		\caption{}
		\label{fig:rod_u1}
	\end{subfigure}
	\begin{subfigure}{0.32\textwidth}
		\includegraphics[width=\linewidth]{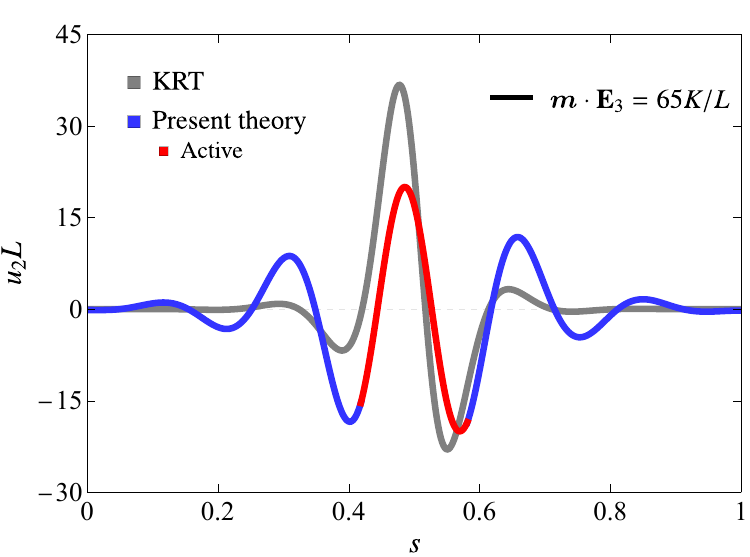}
		\caption{}
		\label{fig:rod_u2}
	\end{subfigure}
	\begin{subfigure}{0.32\textwidth}
		\includegraphics[width=0.98\linewidth]{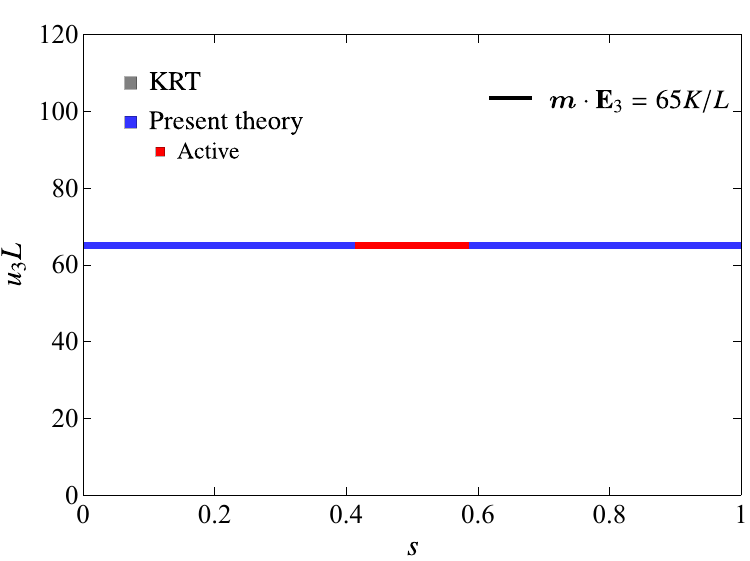}
		\caption{}
		\label{fig:rod_u3}
	\end{subfigure}
    \\
	\begin{subfigure}{0.32\textwidth}
		\includegraphics[width=\linewidth]{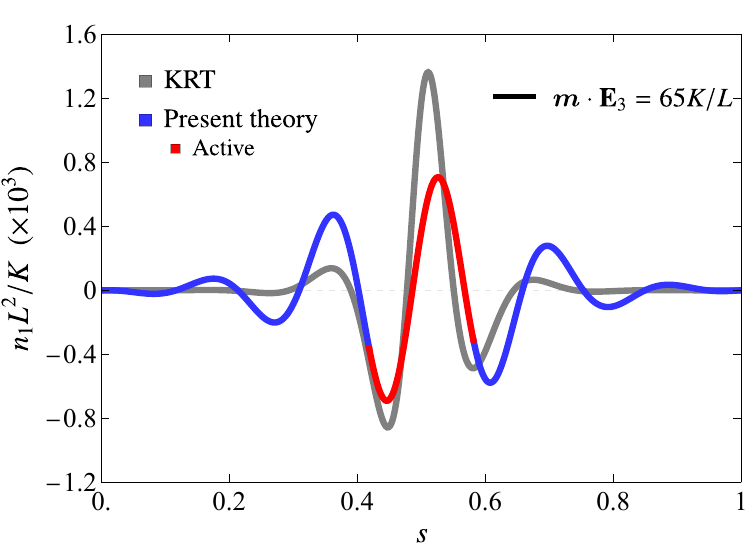}
		\caption{}
		\label{fig:rod_n1}
	\end{subfigure}
	\begin{subfigure}{0.32\textwidth}
		\includegraphics[width=\linewidth]{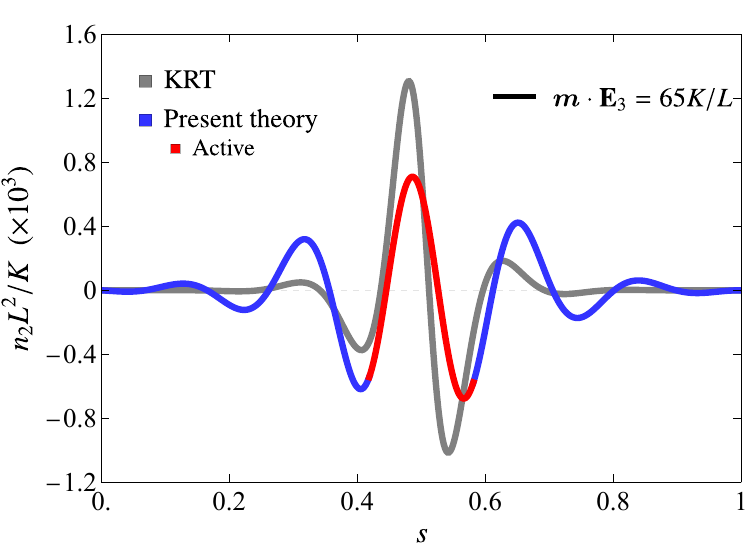}
		\caption{}
		\label{fig:rod_n2}
	\end{subfigure}
	\begin{subfigure}{0.32\textwidth}
		\includegraphics[width=0.98\linewidth]{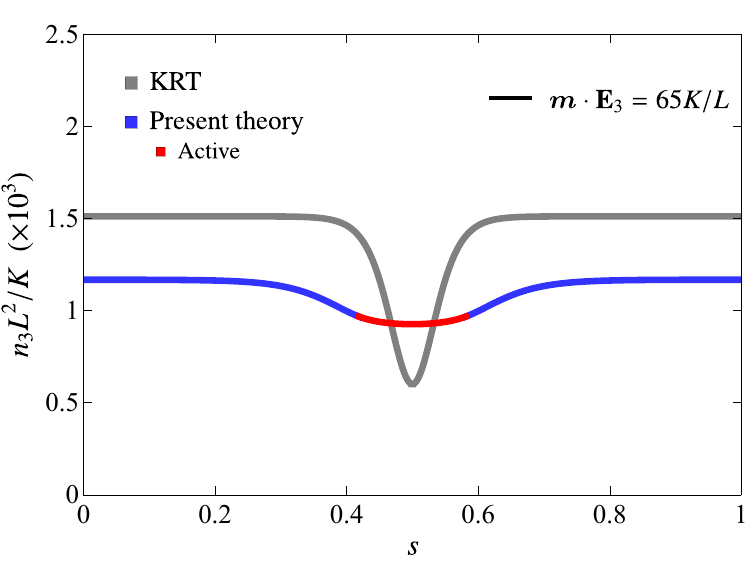}
		\caption{}
		\label{fig:rod_n3}
	\end{subfigure}
	 \\
    \begin{subfigure}{0.32\textwidth}
		\includegraphics[width=\linewidth]{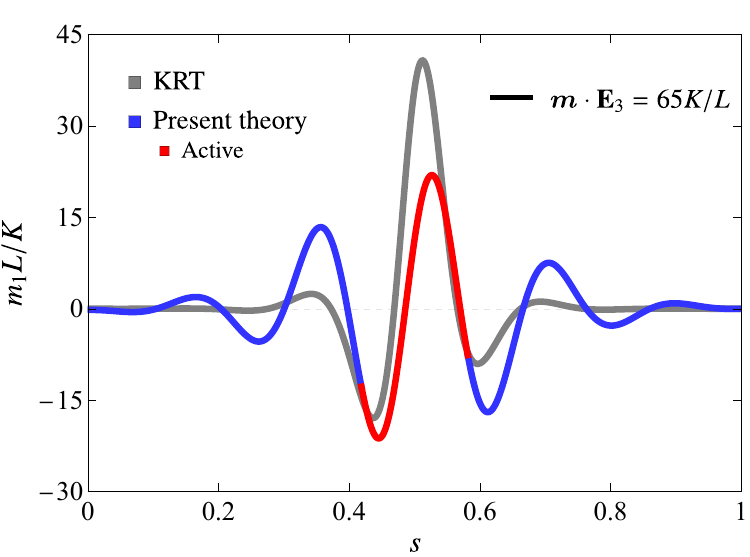}
		\caption{}
		\label{fig:rod_m1}
	\end{subfigure}
	\begin{subfigure}{0.32\textwidth}
		\includegraphics[width=\linewidth]{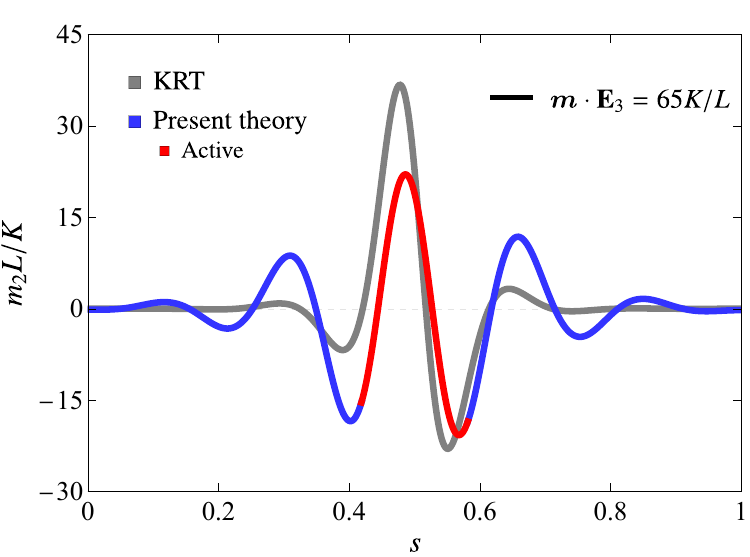}
		\caption{}
		\label{fig:rod_m2}
	\end{subfigure}
	\begin{subfigure}{0.32\textwidth}
		\includegraphics[width=0.98\linewidth]{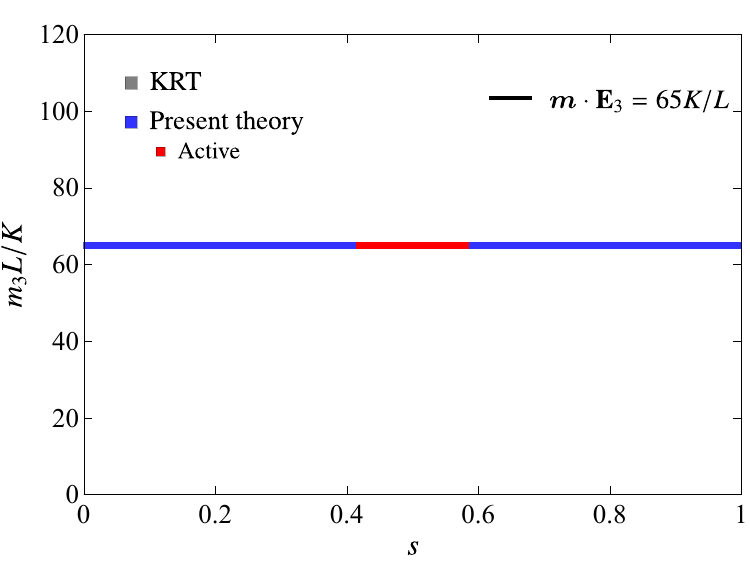}
		\caption{}
		\label{fig:rod_m3}
	\end{subfigure}
	\caption{The top row shows the variations of the normalized bending strains (\ref{fig:rod_u1}) $u_1L$, \ref{fig:rod_u2}) $u_2L$, and (\ref{fig:rod_u3}) $u_3L$ in a Kirchhoff rod and a locally impenetrable elastic tube with an applied moment $\bmo \cdot \bE_3 = 65 K/L$. Panels \ref{fig:rod_n1} to \ref{fig:rod_m3} show the variations in director components of internal forces $\{n_1L^2/K,~n_2L^2/K,~n_3L^2/K\}$ and moments $\{m_1L/K,~m_2L/K,~m_3L/K\}$, respectively. The results for both the Kirchhoff rod (gray curves) and the locally impenetrable elastic tube (blue and red curves) are shown, with red indicating the active region.}
	\label{fig:rod_2}
\end{figure}
The director components of the force and moment balance, obtained from \eqref{eq:force_moment_balance} with $\mathcal{V}=0$, and using the constitutive law in the inactive region \eqref{eq:constitutive_law_inactive} to eliminate $u_i$, are stated as
\begin{equation}\label{eq:3D_force_moment_balance_inactive_scalar}
    \begin{rcases}
        \begin{aligned}
            n'_1 &= - \frac{m_2 n_3}{K} + \frac{m_3 n_2}{K_3}\, ,& n'_2 &= \frac{m_1 n_3}{K} - \frac{m_3 n_1}{K_3}\, ,&n'_3 &= -\frac{m_1 n_2 - m_2 n_1}{K}\, ,\\
            m'_1 &= -\frac{m_2 m_3}{K} + \frac{m_3 m_2}{K_3} + n_2\, ,&m'_2 &= \frac{m_1 m_3}{K} - \frac{m_3 m_1}{K_3} - n_1\, ,& m'_3  &= 0\, ,
        \end{aligned}
    \end{rcases} \forall s \in (0,L) \setminus (s_1,s_2)\, .
\end{equation}
Similarly, the director components of the force and moment balance in the active region are obtained from \eqref{eq:force_moment_balance} with $\mathcal{V}=0$ and \eqref{eq:nonlinear_constitutive_relation_3D} to obtain
\begin{equation}\label{eq:3D_force_moment_balance_active_scalar}
    \begin{rcases}
        \begin{aligned}
            n'_1 &= - \frac{\hat\kappa m_2 n_3}{\left(m_1^2 + m_2^2\right)^{1/2}} + \frac{m_3 n_2}{K_3}\, ,&  n'_2 &= \frac{\hat\kappa m_1 n_3}{\left(m_1^2 + m_2^2\right)^{1/2}} -\frac{m_3 n_1}{K_3}\, ,&n'_3 &= -\frac{\hat\kappa(m_1 n_2 - m_2 n_1)}{\left(m_1^2 + m_2^2\right)^{1/2}}\, ,\\
            m'_1 &= -\frac{\hat\kappa m_2 m_3}{\left(m_1^2 + m_2^2\right)^{1/2}} + \frac{m_3 m_2}{K_3} + n_2\, ,& m'_2 &= \frac{\hat\kappa m_1 m_3}{\left(m_1^2 + m_2^2\right)^{1/2}} - \frac{m_3 m_1}{K_3} - n_1\, ,& m'_3  &= 0\, .
        \end{aligned}
    \end{rcases} \forall s\in(s_1,s_2)\, .
\end{equation}

As in the previous two sections, we divide the length of the rod into three regions, namely $A$, $B$, and $C$, where equations \eqref{eq:3D_force_moment_balance_inactive_scalar} are valid in regions $A$ and $C$, whereas equations \eqref{eq:3D_force_moment_balance_active_scalar} are valid in region $B$.
In total, the full boundary value problem comprises the following unknown vector $\{r^A_i,q^A_\mathsf{j},n^A_i,m^A_i,r^B_i,q^B_\mathsf{j},n^B_i,m^B_i,r^C_i,q^C_\mathsf{j},n^C_i,m^C_i,s_1,s_2\}$ where $i\in\{1,2,3\}$ and $\mathsf{j}\in\{1,2,3,4\}$, which makes it a 41-dimensional problem.

We present a numerical example of an elastic rod with $t=0.05L$ and $K_3/K = 1$, which are the only two relevant parameters that remain in the problem after non-dimensionalizing the system with a length scale $L$ and force scale $K/L^2$. 
The full set of non-dimensionalized equations is stated in Appendix \ref{app:3D_solutions}.

The position and the directors at $s=0$ are constrained as follows,
\begin{equation}\label{eq:boundary_conditions_3D_0}
    r^A_i(0) = 0\, ,\qquad q^A_i(0) = 0\, ,\qquad q^A_4(0) = 1\, ,
\end{equation}
where the conditions on the Euler-parameters ensure that the directors at $s=0$ align with the Cartesian basis.
A twisting moment $M$ is applied along $\bE_3$ at $s=L$ with the tangent constrained to coincide with $\bE_3$ and translations permitted along $\bE_3$ only.
These boundary conditions are stated as
\begin{align}
    r_1^C(L) &= 0\, ,\qquad & r_2^C(L) &= 0\, ,\qquad & \bd_3^C(L)\cdot(\bE_1+\bE_2) &= 0\, ,\label{eq:boundary_conditoin_3D_L_kinematic}\\
    \bn^C(L)\cdot\bE_3 &= 0\, ,\qquad& \bmo^C(L)\cdot\bE_3 &= M\, ,\qquad& \bmo^C(L)\cdot(\bE_1 + \bE_2) &= 0\, .{\label{eq:boundary_conditoin_3D_L_force_moments}}
\end{align}
Finally, relevant jump conditions valid at the boundaries of the active region are given by
\begin{subequations}\label{eq:3D_connecting_conditions}
    \begin{align} 
        \jump{r_i}\big|_{\{s_1,s_2\}} &= 0\, ,&\quad \jump{q_j}\big|_{\{s_1,s_2\}} &= 0\, ,&\quad \jump{n_i}\big|_{\{s_1,s_2\}} &= 0\, ,\\
        \jump{m_i}\big|_{\{s_1,s_2\}} &= 0\, ,&\quad \jump{H}\big|_{\{s_1,s_2\}} &= 0\, .
    \end{align}
\end{subequations}
We solve the problem using a continuation scheme implemented in \texttt{AUTO-07P} \cite{doedel1998}, where our continuation parameter is $M$.
As $M$ is continually increased, a pitchfork bifurcation occurs at $M=8.98L/K$.
We pursue the bifurcating branch and pick a post-buckled configuration at $M=8.50L/K$, where the $s=L$ end of the rod has moved inward.
At this point, we replace the boundary condition \eqref{eq:boundary_conditoin_3D_L_force_moments}$_1$ with $r_3(L) = 0.943L$.
The moment is then continued further until $M=49.28L/K$, which is when the highest Frenet curvature at the middle of the rod reaches its peak value $\hat\kappa=1/0.05L = 20/L$.
Further increase in the moment $M$ leads to the nucleation and spreading of an active region of constant Frenet curvature across the length of the tube.

A comparison between the solution obtained using the Kirchhoff rod theory and the present theory is depicted in Figs.~\ref{fig:rod_config1} and \ref{fig:rod_config2}.
In the former, the Frenet curvature continues to increase and exceeds the upper bound imposed on it by its thickness, and therefore results in configurations in which the material surrounding the centerline curve has locally penetrated.
In the latter, the curvature constraints come into effect and result in the nucleation and propagation of a helix in the active region (shown in red).
A comparison of the Frenet curvature is shown in Fig.~\ref{fig:rod_kappa} and the multiplier $\Lambda$ is shown in Fig.~\ref{fig:rod_lambda}.

Other field quantities for the standard Kirchhoff rod solution and the present theory are shown in Fig.~\ref{fig:rod_2}.
The bending strains are shown in Figs.~\ref{fig:rod_u1}, \ref{fig:rod_u2}, and \ref{fig:rod_u3}.
It is clear that the difference between the Kirchhoff rod solution and the present theory is appreciable even in the inactive regions of the rod.
The twisting strains $u_3$ remain identical to the Kirchhoff rod solution.
This is particularly due to the integral \eqref{eq:twist_conservation}, which remains unaffected by the impenetrability constraint.
The director components of the internal force and moment are shown in Figs.~\ref{fig:rod_n1} to \ref{fig:rod_m3}.
There are significant differences in all these functions both in and outside of the active region, except for $m_3$, which is identical to the Kirchhoff solution for the same reason as for $u_3$.

\section{Conclusion and future work}\label{sec:conclusion}
We derived a theory of elastic tubes with the constraint of local impenetrability actively enforced.
We started with the local impenetrability constraint \eqref{eq:det_F} on the full 3D deformation, and reduced it to an inequality constraint on the Frenet curvature of the base curve (or centerline) of the tube.

Equations governing equilibria of a locally impenetrable elastic tube were derived from the extremization of a functional augmented with the local impenetrability constraint.
The resulting equations revealed that under general loading, the impenetrability constraint activates at a point where the Frenet curvature reaches the upper bound, and spreads into a finite region of constant Frenet curvature (the active region) as the loading progresses.
The extent and location of the active region are unknown \emph{a priori} and are computed as part of the solution.
The active region exhibits a constitutive behavior that is qualitatively different from the inactive regions.
Therefore, a deformed configuration of a locally impenetrable elastic tube can also be thought of as comprising two ``phases'', each with its own constitutive relation.
The jump conditions connecting the two phases are also delivered from the variational principle. 

Our analysis further shows that the standard integrals of Kirchhoff rod theory persist along the full length of the tube, even though the inactive and active regions display qualitatively distinct constitutive behavior.

We illustrated our theory using three examples where standard Kirchhoff theory results in solutions with extremely high curvatures, leading to local penetration of the material of the tube.
The first was a fully flexible tube hanging under self-weight, where the local impenetrability constraint delivered considerably different shapes when compared to standard catenary solutions. 
More interestingly, the active regions in this case lead to the generation of bending moment and shear resistance, which are otherwise absent in the inactive regions.
Next, we showed, using an example of an extremely soft elastic tube hanging under self-weight, that even though bending elasticity may lower the peak curvatures, thus avoiding local penetration, the local impenetrability constraint may become active in certain regimes.

Finally, we considered a highly twisted Kirchhoff rod.
Solutions from the standard KRT result in the concentration of curvature in a small region of the rod.
Solutions generated with the present theory, on the other hand, show the nucleation and propagation of a helix across the length of the tube as the twist was increased. 
These solutions are markedly different, both quantitatively and qualitatively, from the ones delivered by the standard Kirchhoff rod theory.

Extending the present theory to incorporate dynamic effects would be highly relevant to ``chain paradoxes''~\cite{epifanio2015} involving discrete links. Examples include the faster-than-gravity acceleration of a tip falling vertically onto a table~\cite{hammgeminard10,grewal11} or a released folded chain~\cite{calkinmarch89,wongyasui06}, as well as the counterintuitive siphoning of the ``chain fountain''~\cite{mouldsiphon2,epifanio2014,biggins14,martins19}. Literature suggests these behaviors arise in macroscopically constrained chains, such as bead or ball chains, which cannot bend beyond a maximum curvature due to the local impenetrability of their constituents~\cite{hammgeminard10, martins19, biggins14}. Hamm and G{\'{e}}minard~\cite{hammgeminard10} hypothesized that this curvature limit drives the faster-than-gravity tip acceleration, a mechanism confirmed experimentally in~\cite{corbin2018}. Our proposed framework offers a first step toward capturing the complex mechanics of these systems using continuum rod models without introducing extra constitutive parameters~\cite{biggins14,epifanio2014,epifanio2015}.

Another set of problems where the proposed theory could be utilized is in the optimal packing of filaments in a confined volume.
Local impenetrability greatly affects the optimal packing of flexible filaments, as it imposes a hard constraint on the extent to which each tube may bend~\cite{przybyl2001,greenvall2024}.
Questions of optimal packing of filaments also arise in the context of biomolecules, such as DNA, and have been widely studied under the purview of rod models~\cite{benham1977,banavar2007,maritan2000,smith2008,sogo1999,stasiak2000,westcott1997,goyal2008,katritch1996,purohit2008,schlick1995,travers2004}.
While a substantial part of the literature on the subject~\cite{balmaz2012,chamekh2009,coleman2000,starostin2004,starostin2014}, partly inspired by its relevance to biological structural shapes, accounts for global impenetrability, the constraint of local impenetrability has received far less attention outside the realm of pure mathematics~\cite{gonzalez2002,hoffman2022}.

The ultimate extension of our theory lies in the incorporation of global contact, which would pave the way for a complete theory of self-avoiding elastic rods. While we do not anticipate this extension to be straightforward, we believe that it is certainly worth pursuing.

\section{Acknowledgements}
HS gratefully acknowledges financial support for this work by the Science Engineering and Research Board (SERB) through grant SRG/2023/000079.

\appendix

\section{Variational derivation}\label{app:variational_derivation}
The variation in the action \eqref{eq:action_functional} induced by \eqref{eq:variations}, i.e. $\delta\mathcal{A}\equiv\mathcal{A}^* - \mathcal{A}$ is written as
\begin{align}\label{eq:variation_energy_0}
    \delta\mathcal{A} = \int_{0^*}^{L^*}\mathcal{L}\left(\br^*(s^*),(\br^*(s^*))',\bd^*_i(s^*),u^*_i(s^*),\kappa^*(s^*),g^*(s^*)\right)ds^* -  \int_0^L \mathcal{L}\left(\br(s),\br'(s),\bd_i(s),u_i(s),\kappa(s),g(s)\right)ds\, .
\end{align}
Due to the variation in $s$, the domain of the first integral does not coincide with the second one.
To remedy this situation, we transform the domain of the first integral from $[0^*, L^*]\rightarrow[0,L]$, via \eqref{eq:variations} $ds^* = (\partial s^*/\partial s) ds = (1+\partial\delta s/\partial s) ds$.
Consequently, \eqref{eq:variation_energy_0} can be re-arranged as
\begin{align}\label{eq:variation_energy_1}
    \delta\mathcal{A} = \int_0^L\mathcal{L}\frac{\partial\delta s}{\partial s} ds + \int_0^L \left\{\mathcal{L}(\br+\delta\br,\br'+\delta(\br'),\bd_i + \delta\bd_i,u_i+\delta u_i, \kappa+\delta\kappa, g+\delta g) - \mathcal{L}(\br,\br',\bd_i,u_i,\kappa,g)\right\} ds\, .
\end{align}
The first order difference in the Lagrangian densities in the second integral can be written as 
\begin{align}\label{eq:variation_energy_2}
    \delta\mathcal{A} = \int_0^L\mathcal{L}\frac{\partial\delta s}{\partial s} ds + \int_0^L\left\{\frac{\partial\mathcal{L}}{\partial\br}\cdot\delta\br + \frac{\partial\mathcal{L}}{\partial\br'}\cdot\delta(\br') + \frac{\partial\mathcal{L}}{\partial\bd_i}\cdot\delta\bd_i + 
    \frac{\partial\mathcal{L}}{\partial u_i}\delta u_i+
    \frac{\partial\mathcal{L}}{\partial \kappa}\delta \kappa+ \frac{\partial\mathcal{L}}{\partial g}\delta g\right\}ds\,.
\end{align}
The definition \eqref{eq:variations} of the variations $\delta$ in various field quantities is such that the concerned fields on the two sides of the equations are functions of different arguments \cite{hill1951}.
A major consequence of this fact is that the operator $\delta$ and derivative $()'$ w.r.t. $s$ do not commute.
To remedy this problem, we define an operator $\bar\delta$ which represents the variation of a field at the same material label $s$, and therefore commutes with derivatie w.r.t. s.
This is related to $\delta$ as,
\begin{align}\label{eq:variations_2}
    \delta\br(s)=\bar\delta\br(s) + \br'(s)\delta s\, ,\quad\delta\bd_i(s) = \bar\delta\bd_i(s)+\bd'_i(s)\delta s\, ,\quad \delta g(s) = \bar\delta g(s) + g'(s)\delta s\,.
\end{align}
Using the above in \eqref{eq:variation_energy_2} we obtain the following
\begin{align}
    \delta\mathcal{A} = \int_0^L\mathcal{L}\frac{\partial\delta s}{\partial s} ds &+ \int_0^L\left\{\frac{\partial\mathcal{L}}{\partial\br}\cdot\br' + \frac{\partial\mathcal{L}}{\partial\br'}\cdot\br'' + \frac{\partial\mathcal{L}}{\partial\bd_i}\cdot\bd'_i + \frac{\partial\mathcal{L}}{\partial u_i}u'_i + \frac{\partial\mathcal{L}}{\partial\kappa}\kappa' + \frac{\partial\mathcal{L}}{\partial g}g'\right\}\delta s\, ds \nonumber\\
    & + \int_0^L\left\{\frac{\partial\mathcal{L}}{\partial\br}\cdot\bar\delta\br + \frac{\partial\mathcal{L}}{\partial\br'}\cdot\bar\delta(\br') + \frac{\partial\mathcal{L}}{\partial\bd_i}\cdot\bar\delta\bd_i + 
    \frac{\partial\mathcal{L}}{\partial u_i}\bar\delta u_i+
    \frac{\partial\mathcal{L}}{\partial \kappa}\bar\delta \kappa+ \frac{\partial\mathcal{L}}{\partial g}\bar\delta g\right\}ds\, .
\end{align}
The second integral on the right side can be identified as the total derivative of the Lagrangian function,
\begin{align}
    \delta\mathcal{A} = &\int_0^L\mathcal{L}\frac{\partial\delta s}{\partial s} ds +\int_0^L \frac{d\mathcal{L}}{ds}\delta s\, ds + \int_0^L\left\{\frac{\partial\mathcal{L}}{\partial\br}\cdot\bar\delta\br + \frac{\partial\mathcal{L}}{\partial\br'}\cdot\bar\delta(\br') + \frac{\partial\mathcal{L}}{\partial\bd_i}\cdot\bar\delta\bd_i + 
    \frac{\partial\mathcal{L}}{\partial u_i}\bar\delta u_i+
    \frac{\partial\mathcal{L}}{\partial \kappa}\bar\delta \kappa+ \frac{\partial\mathcal{L}}{\partial g}\bar\delta g\right\}ds\, .
\end{align}
The integrands of the first two integrals combine as $d(\mathcal{L}\delta s)/ds$,
\begin{align}\label{eq:variation_energy_3}
    \delta\mathcal{A} = &\int_0^L\frac{d}{ds}\left(\mathcal{L}\delta s\right) ds + \int_0^L\left\{\frac{\partial\mathcal{L}}{\partial\br}\cdot\bar\delta\br + \frac{\partial\mathcal{L}}{\partial\br'}\cdot\bar\delta(\br') + \frac{\partial\mathcal{L}}{\partial\bd_i}\cdot\bar\delta\bd_i + 
    \frac{\partial\mathcal{L}}{\partial u_i}\bar\delta u_i+
    \frac{\partial\mathcal{L}}{\partial \kappa}\bar\delta \kappa+ \frac{\partial\mathcal{L}}{\partial g}\bar\delta g\right\}ds\, .
\end{align}
Variations $\bar\delta\bd_i$, $\bar\delta u_i$, and $\bar\delta\kappa$ can be computed as,
\begin{align}\label{eq:variations_d_u_kappa}
    \bar\delta\bd_i = \bar\delta{\bm \theta}\times\bd_i\, ,\qquad  \bar\delta u_i = \bar\delta\bm{\theta}'\cdot\bd_i\, ,\qquad\bar\delta\kappa = \frac{1}{\kappa}\mathbb{P}\bu\cdot\bar\delta\bm{\theta}'\, .
\end{align}
Note that $\bar\delta u_i = \bar\delta (\bu\cdot\bd_i)$.
The first form ensures that the perturbations in the director frame preserve orthonormality, whereas the second is a consequence of the relation $\kappa=\sqrt{u_1^2+u_2^2}$.
Here $\mathbb{P} = \bI - \bd_3\otimes\bd_3$ is a projection operator that projects any vector onto the cross-sectional plane spanned by $\{\bd_1,\bd_2\}$.
Using \eqref{eq:variations_d_u_kappa} into \eqref{eq:variation_energy_3}, and using standard procedure to integrate by parts, we obtain the following upon some rearrangement
\begin{align}\label{eq:variation_4}
    \delta\mathcal{A} = &\int_0^L\left[\frac{\partial\mathcal{L}}{\partial\br'}\cdot\delta\br + \left(\frac{\partial\mathcal{L}}{\partial u_i}\bd_i + \frac{\partial\mathcal{L}}{\partial\kappa}\frac{1}{\kappa}\mathbb{P}\bu\right)\cdot\delta\bm{\theta} + \left\{\mathcal{L}-\frac{\partial\mathcal{L}}{\partial\br'}\cdot\br' - \left(\frac{\partial\mathcal{L}}{\partial u_i}\bd_i + \frac{\partial\mathcal{L}}{\partial\kappa}\frac{1}{\kappa}\mathbb{P}\bu\right)\cdot\bu\right\}\delta s\right]'\, ,\\
    &-\int_0^L\left[\left(\frac{\partial\mathcal{L}}{\partial\br'}\right)'-\frac{\partial\mathcal{L}}{\partial\br}\right]\cdot\delta\br ds - \int_0^L \left[\left(\frac{\partial\mathcal{L}}{\partial u_i}\bd_i + \frac{\partial\mathcal{L}}{\partial\kappa}\frac{1}{\kappa}\mathbb{P}\bu\right)' - \bd_i\times\frac{\partial\mathcal{L}}{\partial\bd_i}\right]\cdot\delta\bm{\theta} ds\\
    & + \int_0^L\left\{\left[\left(\frac{\partial\mathcal{L}}{\partial\br'}\right)'-\frac{\partial\mathcal{L}}{\partial\br}\right]\cdot\br' + 
    \left[\left(\frac{\partial\mathcal{L}}{\partial u_i}\bd_i + \frac{\partial\mathcal{L}}{\partial\kappa}\frac{1}{\kappa}\mathbb{P}\bu\right)' - \bd_i\times\frac{\partial\mathcal{L}}{\partial\bd_i}\right]\cdot\bu - \frac{\partial\mathcal{L}}{\partial g}g'
    \right\}\delta s ds  + \int_0^L\frac{\partial\mathcal{L}}{\partial g}\delta g ds
\end{align}
where $\bar\delta$ has been eliminated in favor of $\delta$ using \eqref{eq:variations_2}.

\section{BVP for an elastic tube hanging under self-weight}\label{app:bvp_elastic_tube}
The full non-dimensional boundary value problem for an elastic tube hanging under gravity is stated as follows, 
\begin{subequations}\label{eq:bvp_elastica}
	\begin{align}
		\frac{dr_1^A}{ds^A}&= \sin \theta^A\,,\qquad& r_1^A(0)&=0, \label{eq:bvp_elastica_A_r1}\\
		\frac{dr_3^A}{ds^A}&= \cos \theta^A\,,\qquad& r_3^A(0)&=0, \label{eq:bvp_elastica_A_r3}\\
		\frac{d\theta^A}{ds^A}&= {m_2}^A\,,\qquad& m_2^A(0)&=0, 
        \label{eq:bvp_elastica_A_theta}\\
		\frac{dn_1^A}{ds^A}&= -n_3^Am_2^A + \alpha \cos \theta^A \,,\qquad& r_1^C(1)&=0, \label{eq:bvp_elastica_A_n1}\\
		\frac{dn_3^A}{ds^A}&=  n_1^A m_2^A+ \alpha \sin \theta^A\,,\qquad& r_3^C(1)&=\Delta, 	\label{eq:bvp_elastica_A_n3}\\
		\frac{dm_2^A}{ds^A}&= {-n_1^A}\,,\qquad& m_2^C(1)&=0, \label{eq:bvp_elastica_A_m2} \\
		\frac{dr_1^B}{ds^B}&= \sin \theta^B\,,\qquad& r_1^A(s_1)&=r_1^B(s_1), \label{eq:bvp_elastica_B_r1} \\
		\frac{dr_3^B}{ds^B}&= \cos \theta^B\,,\qquad& r_3^A(s_1)&=r_3^B(s_1), \label{eq:bvp_elastica_B_r3} \\
		\frac{d\theta^B}{ds^B}&= \hat\kappa \,,\qquad& \theta^A(s_1)&=\theta^B(s_1), \label{eq:bvp_elastica_B_theta} \\
		\frac{dn_1^B}{ds^B}&=- n_3^B \hat\kappa+ \alpha \cos \theta^B \,,\qquad& n_1^A(s_1)&=n_1^B(s_1), \label{eq:bvp_elastica_B_n1} \\
		\frac{dn_3^B}{ds^B}&= n_1^B \hat\kappa+ \alpha \sin \theta^B\,,\qquad& n_3^A(s_1)&=n_3^B(s_1), \label{eq:bvp_elastica_B_n3} \\
		\frac{dm_2^B}{ds^B}&=-n_1^B 	\, ,\qquad&  m_2^A(s_1)&=m_2^B(s_1), \label{eq:bvp_elastica_B_m2} \\
		\frac{dr_1^C}{ds^C}&= \sin \theta^C\,,\qquad& r_1^B(s_2)&=r_1^C(s_2) , \label{eq:bvp_elastica_C_r1} \\
		\frac{dr_3^C}{ds^C}&= \cos \theta^C\,,\qquad& r_3^B(s_2)&=r_3^C(s_2) , \label{eq:bvp_elastica_C_r3} \\
		\frac{d\theta^C}{ds^C}&= {u_2}^C\,,\qquad& \theta^B(s_2)&=\theta^C(s_2), \label{eq:bvp_elastica_C_theta}\\
		\frac{dn_1^C}{ds^C}&= - n_3^C m_2^C+ \alpha \cos \theta^C \,,\qquad& n_1^B(s_2)&=n_1^C(s_2), \label{eq:bvp_elastica_C_n1}  \\
		\frac{dn_3^C}{ds^C}&=  n_1^C m_2^C+ \alpha\sin \theta^C\,,\qquad& n_3^B(s_2)&=n_3^C(s_2),  \label{eq:bvp_elastica_C_n3} \\
		\frac{dm_2^C}{ds^C}&= {-n_1^C}\,,\qquad& m_2^B(s_2)&=m_2^C(s_2), \label{eq:bvp_elastica_C_m2}\\
		\frac{ds_1}{ds^A}&=0\,,\qquad & n_3^A(s_1)+\frac{1}{2} \left(m_2^A(s_1)\right)^2&= n_3^B(s_1)+m_2^B(s_1) \hat\kappa-\frac{1}{2}  \hat\kappa^2 , \label{eq:bvp_elastica_s1} \\
		\frac{ds_2}{ds^C}&=0\,,\qquad &  n_3^B(s_2)+ m_2^B(s_2) \hat\kappa-\frac{1}{2} \hat\kappa^2&= n_3^C(s_2)+\frac{1}{2} \left(m_2^C(s_2)\right)^2. \label{eq:bvp_elastica_s2}
	\end{align}
\end{subequations}
The system has been non-dimensionalized by a length scale $L$ and a force scale $K/L^2$.
The constant $\alpha= \rho g L^3/K$ appear as the governing parameter of the problem. 
Equations \ref{eq:bvp_elastica_A_r1} to \ref{eq:bvp_elastica_A_m2}  and \ref{eq:bvp_elastica_C_r1} to \ref{eq:bvp_elastica_C_m2} are the kinematics and force balance equations for the inactive regions $A$ and $C$, respectively. Similarly, equations  \ref{eq:bvp_elastica_B_r1} to \ref{eq:bvp_elastica_B_m2} correspond to the kinematics and force balance equations for the active region $B$.
Equations \ref{eq:bvp_elastica_s1} and \ref{eq:bvp_elastica_s2} are trivial equations indicating that $s_1$ and $s_2$ as constant functions to be determined as part of the solution. 
Before feeding the boundary value problem to AUTO-07P, the domains of regions $A$, $B$, and $C$ are mapped to $\xi \in [0,1]$ as $s^A=s_1 \xi$, $s^B=(s_2-s_1) \xi+s_1$ and $s^C=(1-s_2) \xi+s_2$, in order to facilitate numerical implementation.
\section{BVP for 3D twisted solutions}\label{app:3D_solutions}
The full non-dimensional  boundary value problem for the  3D solutions of an elastic tube with the lengths scaled by $L$ and forces by $K/L^2$ is given as,
\begin{subequations} \label{eq:3D_elastic_tube}
    \begin{align}
    	\frac{dr_1^A}{ds^A}&= 2 (q_1^A q_3^A +  q_2^A q_4^A)\,,\qquad& r_1^A(0)&=0, \label{eq:3D_elastic_tube_A_r1}\\
		\frac{dr_2^A}{ds^A}&= 2 (q_2^A q_3^A -  q_1^A q_4^A)\,,\qquad& r_2^A(0)&=0, \label{eq:3D_elastic_tube_A_r2}\\
		\frac{dr_3^A}{ds^A}&=(-{q_1^A}^2-{q_2^A}^2+{q_3^A}^2+{q_4^A}^2)\,,\qquad& r_3^A(0)&=0, \label{eq:3D_elastic_tube_A_r3}\\
        \frac{dq_1^A}{ds^A} &= \frac{1}{2}\left( \frac{q_4^Am_1^A -  q_3^Am_2^A}{K} + \frac{ q_2^Am_3^A}{K_3}\right)\,,\qquad& q_1^A(0)&=0, \label{eq:3D_elastic_tube_A_q1}\\
		\frac{dq_2^A}{ds^A} &= \frac{1}{2}\left( \frac{q_3^A m_1^A + q_4^A m_2^A}{K} - \frac{q_1^Am_3^A}{K_3}\right)\,,\qquad& q_2^A(0)&=0, \label{eq:3D_elastic_tube_A_q2} \\
		\frac{dq_3^A}{ds^A} &= \frac{1}{2}\left(- \frac{q_2^Am_1^A-  q_1^Am_2^A}{K} +  \frac{q_4^Am_3^A}{K_3}\right)\,,\qquad& q_3^A(0)&=0, \label{eq:3D_elastic_tube_A_q3}\\
		\frac{dq_4^A}{ds^A} &= \frac{1}{2}\left(- \frac{q_1^Am_1^A +  q_2^Am_2^A}{K} -  \frac{q_3^Am_3^A}{K_3}\right)\,,\qquad& q_4^A(0)&=1, \label{eq:3D_elastic_tube_A_q4}\\
        \frac{dn_1^A}{ds^A} &=- \frac{m_2^A n_3^A}{K} + \frac{m_3^A n_2^A}{K_3}\,,\qquad& r_1^A(s_1)&=r_1^B(s_1), \label{eq:3D_elastic_tube_A_n1} \\
		\frac{dn_2^A}{ds^A} &=\frac{m_1^A n_3^A}{K} - \frac{m_3^A n_1^A}{K_3} \,,\qquad& r_2^A(s_1)&=r_2^B(s_1), \label{eq:3D_elastic_tube_A_n2}\\
		\frac{dn_3^A}{ds^A} &= - \frac{m_1^An_2^A- m_2^An_1^A}{K}\,,\qquad& r_3^A(s_1)&=r_3^B(s_1), \label{eq:3D_elastic_tube_A_n3}\\
		\frac{dm_1^A}{ds^A} &=-\frac{m_2^A m_3^A}{K} + \frac{m_3^A m_2^A}{K_3} + n_2^A\,,\qquad& q_1^A(s_1)&=\ q_1^B(s_1), \label{eq:3D_elastic_tube_A_m1}\\
		\frac{dm_2^A}{ds^A} &=  \frac{m_1^A m_3^A}{K} - \frac{m_3^A m_1^A}{K_3}-n_1^A\,,\qquad& q_2^A(s_1)&=\ q_2^B(s_1), \label{eq:3D_elastic_tube_A_m2}\\
		\frac{dm_3^A}{ds^A} &= 0\,,\qquad& q_3^A(s_1)&=\ q_3^B(s_1), \label{eq:3D_elastic_tube_A_m3}\\
        \frac{dr_1^B}{ds^B}&= 2 (q_1^B q_3^B +  q_2^B q_4^B)\,,\qquad& 
		q_4^A(s_1)&=\ q_4^B(s_1), \label{eq:3D_elastic_tube_B_r1} \\
		\frac{dr_2^B}{ds^B}&= 2 (q_2^B q_3^B -  q_1^B q_4^B)\,,\qquad&
		n_1^A(s_1)&=n_1^B(s_1),\label{eq:3D_elastic_tube_B_r2}\\
		\frac{dr_3^B}{ds^B}&=(-{q_1^B}^2-{q_2^B}^2+{q_3^B}^2+{q_4^B}^2)\,,\qquad&
		n_2^A(s_1)&=n_2^B(s_1), \label{eq:3D_elastic_tube_B_r3}\\
        \frac{dq_1^B}{ds^B} &= \frac{1}{2}\left( \frac{(q_4^B  m_1^B- q_3^B m_2^B) \hat\kappa}{({m_1^B}^2+{m_2^B}^2)^{1/2}}  +  \frac{q_2^B m_3^B}{K_3}\right)\,,\qquad& 
		n_3^A(s_1)&=n_3^B(s_1), \label{eq:3D_elastic_tube_B_q1}\\
		\frac{dq_2^B}{ds^B} &= \frac{1}{2}\left(\frac{(q_3^B  m_1^B+q_4^B m_2^B)\hat\kappa}{({m_1^B}^2+{m_2^B}^2)^{1/2}} -  \frac{q_1^B m_3^B}{K_3}\right)\,,\qquad& 
		m_1^A(s_1)&=m_1^B(s_1),\label{eq:3D_elastic_tube_B_q2}\\
		\frac{dq_3^B}{ds^B} &= \frac{1}{2}\left(- \frac{(q_2^B m_1^B-q_1^B m_2^B)\hat\kappa}{({m_1^B}^2+{m_2^B}^2)^{1/2}}+  \frac{q_4^B m_3^B}{K_3}\right)\,,\qquad& 
		m_2^A(s_1)&=m_2^B(s_1),\label{eq:3D_elastic_tube_B_q3}\\
		\frac{dq_4^B}{ds^B} &= \frac{1}{2}\left(- \frac{(q_1^B m_1^B+ q_2^B m_2^B)\hat\kappa}{({m_1^B}^2+{m_2^B}^2)^{1/2}} -  \frac{q_3^B m_3^B}{K_3}\right)\,,\qquad&   
		m_3^A(s_1)&=m_3^B(s_1), \label{eq:3D_elastic_tube_B_q4}\\
        \frac{dn_1^B}{ds^B} &= - \frac{\hat\kappa m_2^B n_3^B}{\left({m_1^B}^2 + {m_2^B}^2\right)^{1/2}} + \frac{m_3^B n_2^B}{K_3}\,,\qquad& r_1^B(s_2)&=r_1^C(s_2), \ \label{eq:3D_elastic_tube_B_n1} \displaybreak[1] \\ 
        \frac{dn_2^B}{ds^B} &= \frac{\hat\kappa m_1^B n_3^B}{\left({m_1^B}^2 + {m_2^B}^2\right)^{1/2}} - \frac{m_3^B n_1^B}{K_3}\,,\qquad& r_2^B(s_2)&=r_2^C(s_2), \label{eq:3D_elastic_tube_B_n2} \\
		\frac{dn_3^B}{ds^B} &= -\frac{\hat\kappa \left(m_1^B n_2^B-m_2^B n_1^B\right)}{\left({m_1^B}^2 + {m_2^B}^2\right)^{1/2}}\,,\qquad& 
		r_3^B(s_2)&=r_3^C(s_2),\label{eq:3D_elastic_tube_B_n3} \\
        \frac{dm_1^B}{ds^B} &=- \frac{\hat\kappa m_2^B  m_3^B}{\left({m_1^B}^2 + {m_2^B}^2\right)^{1/2}}+ \frac{m_3^B m_2^B}{K_3}+n_2^B\,,\qquad& 
		q_1^B(s_2)&=q_1^C(s_2), \label{eq:3D_elastic_tube_B_m1}\\
		\frac{dm_2^B}{ds^B} &= \frac{\hat\kappa m_1^B m_3^B }{\left({m_1^B}^2 + {m_2^B}^2\right)^{1/2}}-\frac{m_3^B m_1^B}{K_3}-n_1^B\,,\qquad&  
		q_2^B(s_2)&=q_2^C(s_2), \label{eq:3D_elastic_tube_B_m2}\\
		\frac{dm_3^B}{ds^B} &= 0\,,\qquad& 
		q_3^B(s_2)&=q_3^C(s_2),\label{eq:3D_elastic_tube_B_m3}\\
        \frac{dr_1^C}{ds^C}&= 2 (q_1^C q_3^C +  q_2^C q_4^C)\,,\qquad& q_4^B(s_2)&=q_4^C(s_2), \label{eq:3D_elastic_tube_C_r1}\\
		\frac{dr_2^C}{ds^C}&= 2 (q_2^C q_3^C -  q_1^C q_4^C)\,,\qquad&  n_1^B(s_2)&=n_1^C(s_2), \label{eq:3D_elastic_tube_C_r2}\\
		\frac{dr_3^C}{ds^C}&=(-{q_1^C}^2-{q_2^C}^2+{q_3^C}^2+{q_4^C}^2)\,,\qquad&  n_2^B(s_2)&=n_2^C(s_2), \label{eq:3D_elastic_tube_C_r3}\\
        \frac{dq_1^C}{ds^C} &= \frac{1}{2}\left( \frac{q_4^Cm_1^C -  q_3^Cm_2^C}{K} + \frac{ q_2^Cm_3^C}{K_3}\right)\,,\qquad& n_3^B(s_2)&=n_3^C(s_2), \label{eq:3D_elastic_tube_C_q1}\\
		\frac{dq_2^C}{ds^C} &= \frac{1}{2}\left( \frac{q_3^C m_1^C + q_4^C m_2^C}{K} - \frac{q_1^Cm_3^C}{K_3}\right)\,,\qquad& m_1^B(s_2)&=m_1^C(s_2), \label{eq:3D_elastic_tube_C_q2} \\
		\frac{dq_3^C}{ds^C} &= \frac{1}{2}\left(- \frac{q_2^Cm_1^C-  q_1^Cm_2^C}{K} +  \frac{q_4^Cm_3^C}{K_3}\right)\,,\qquad& m_2^B(s_2)&=m_2^C(s_2), \label{eq:3D_elastic_tube_C_q3}\\
		\frac{dq_4^C}{ds^C} &= \frac{1}{2}\left(- \frac{q_1^Cm_1^C +  q_2^Cm_2^C}{K} -  \frac{q_3^Cm_3^C}{K_3}\right)\,,\qquad& m_3^B(s_2)&=m_3^C(s_2), \label{eq:3D_elastic_tube_C_q4}\\
        \frac{dn_1^C}{ds^C} &=- \frac{m_2^C n_3^C}{K} + \frac{m_3^C n_2^C}{K_3}\,,\qquad&  r_1^C(1)&=0, \label{eq:3D_elastic_tube_C_n1} \\
		\frac{dn_2^C}{ds^C} &=\frac{m_1^C n_3^C}{K} - \frac{m_3^C n_1^C}{K_3} \,,\qquad& r_2^C(1)&=0, \label{eq:3D_elastic_tube_C_n2}\\
		\frac{dn_3^C}{ds^C} &= - \frac{m_1^Cn_2^C- m_2^Cn_1^C}{K}\,,\qquad& \bd_3^C(1) \cdot (\bE_1+\bE_2)&=0, \label{eq:3D_elastic_tube_C_n3}\\
		\frac{dm_1^C}{ds^C} &=-\frac{m_2^C m_3^C}{K} + \frac{m_3^C m_2^C}{K_3} + n_2^C\,,\qquad& \bn^C(1) \cdot \bE_3&=0, \label{eq:3D_elastic_tube_A_m1}\\
		\frac{dm_2^C}{ds^C} &=  \frac{m_1^C m_3^C}{K} - \frac{m_3^C m_1^C}{K_3}-n_1^C\,,\qquad& \bmo^C(1)\cdot \bE_3&=M, \label{eq:3D_elastic_tube_C_m2}\\
		\frac{dm_3^C}{ds^C} &= 0\,,\qquad& \bmo^C(1)\cdot (\bE_1+\bE_2)&=0, \label{eq:3D_elastic_tube_C_m3}\\
        \frac{ds_1}{ds^A}&=0\,,\qquad & H^A(s_1)&=H^B(s_1) , \label{eq:3D_elastic_tube_C_s1} \\
		\frac{ds_2}{ds^C}&=0\,,\qquad & H^B(s_2)&=H^C(s_2)\label{eq:3D_elastic_tube_C_s2}.
    \end{align}    
\end{subequations}
Equations \ref{eq:3D_elastic_tube_A_r1} to \ref{eq:3D_elastic_tube_A_m3}, \ref{eq:3D_elastic_tube_B_r1} to \ref{eq:3D_elastic_tube_B_m3}, and \ref{eq:3D_elastic_tube_C_r1} to \ref{eq:3D_elastic_tube_C_m3}  are the kinematics and force balance equations for the  regions $A$, $B$ and $C$, respectively. The unknown parameters $s_1,s_2$ are treated as constant functions to be determined as part of the solution through trivial equations \ref{eq:3D_elastic_tube_C_s1} and \ref{eq:3D_elastic_tube_C_s2}. The Hamiltonian $H$ for various regions are as follows:
\begin{align}\label{eq:3D_elastic_tube_H}
  H^A&= n_3^A+ \frac{1}{2}\left(\frac{{m_i^A}^2}{K_i}\right)\,,  &  H^B&= n_3^B+ \frac{1}{2}\left(\frac{{m_3^A}^2}{K_3}-K \hat\kappa^2\right)+\left({m_1^B}^2 + {m_2^B}^2\right)^{1/2} \hat\kappa \,,  &  H^C&= n_3^C+ \frac{1}{2}\left(\frac{{m_i^C}^2}{K_i}\right)\,,
\end{align}
where $i \in \{1, 2, 3\}$.
As in Appendix \ref{app:bvp_elastic_tube}, we map various arclengths to a single parameter $\xi \in [0,1]$ as follows:  $s^A=s_1 \xi$, $s^B=(s_2-s_1) \xi+s_1$ and $s^C=(1-s_2) \xi+s_2$, for easier numerical implementation in \texttt{AUTO-07P} \cite{doedel1998}.

%


\end{document}